\documentclass{article}
\usepackage{graphicx}
\usepackage[english]{babel}

\title{
\includegraphics[width=0.35\textwidth]{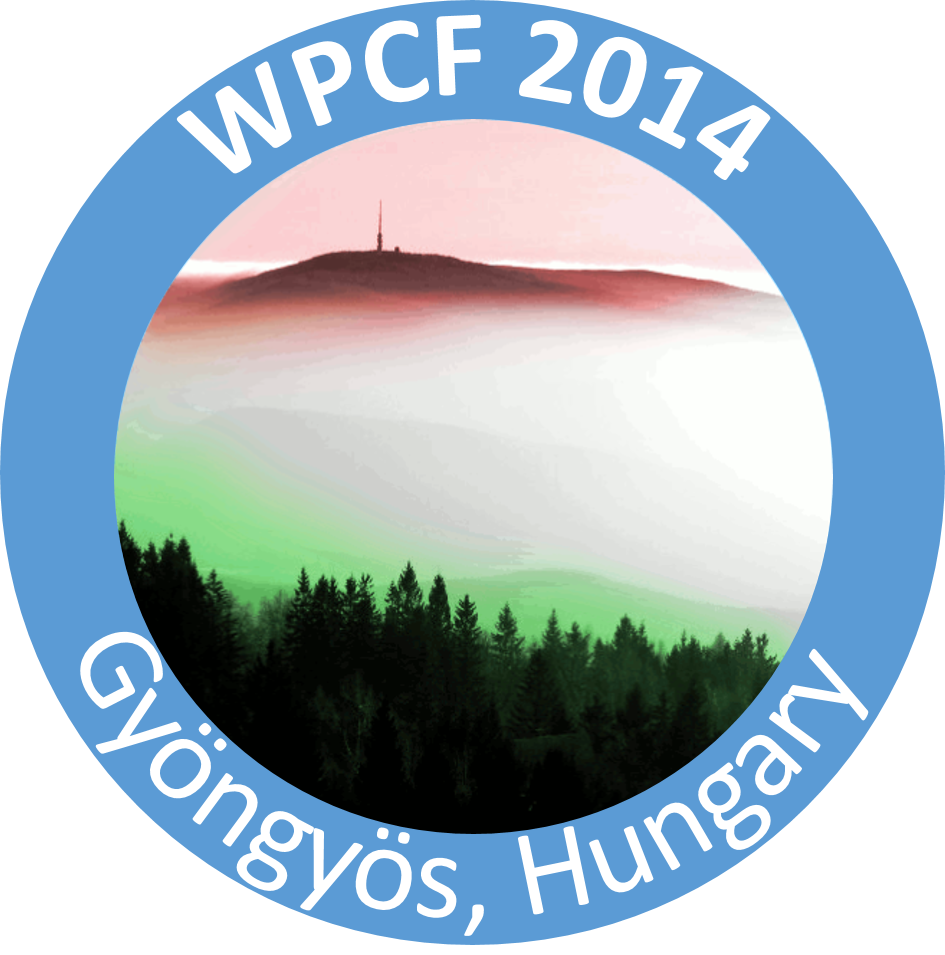}\\[1cm]
Multidimensional analysis of Bose-Einstein correlations in pp collisions at 2.76 and 7 TeV in CMS}
\author{{Sandra S. Padula$^1$  (for the CMS Collaboration)}\\[1ex]
$^1$ Instituto de F\'\i sica Te\'orica--UNESP, S\~ao Paulo, SP, Brazil\\
}

\begin{document}

\fontfamily{lmss}\selectfont
\maketitle

\begin{abstract}
\noindent
Multidimensional two-particle Bose-Einstein correlation functions of charged hadrons are reported for $pp$ collisions at 2.76 and 7 TeV in terms of different components of the pair relative momentum, extending the previous one-dimensional (1-D) analyses of CMS. This allows for investigating the extension of the source accessible to the femtoscopic correlation technique in different directions, revealing a more detailed picture of the emitting source in these collisions at increasing energies. The measurements are performed for different intervals of the pair average transverse momentum, $k_T$, and for increasing charged particle multiplicitiy, $N_{ch}$. Results in 1-D, 2-D and 3-D show a decrease of the fit radius parameters with $k_T$, whereas a clear rise with $N_{ch}$ is observed in all cases. In addition, the fit radius parameters at both energies show close similarity in size and behavior within the same intervals of ($N_{ch},k_T$). 
\end{abstract}

\section{Introduction}
\label{sec:intro}

Femtoscopic Bose-Einstein correlations, also known as HBT/GGLP effect, were investigated in Ref. \cite{cms-hbt-1st,cms-hbt-2nd} by CMS for pp collisions at $\sqrt{s} = $0.9~TeV \cite{cms-hbt-1st,cms-hbt-2nd}, 2.36~TeV \cite{cms-hbt-1st} and 7~TeV \cite{cms-hbt-2nd}. Such phenomenon was discovered by Goldhaber, S. Goldhaber, W. Lee and A. Pais (GGLP effect) \cite{gglp}, being the analogous in high-energy collisions to a similar method proposed by R. Hanbury-Brown and R. Q. Twiss (HBT effect) \cite{hbt} for estimating angular dimensions of stars. A broad investigation was carried out in these studies in terms of the invariant relative momenum $Q_{\textrm{inv}} $.   
In Ref. \cite{cms-hbt-2nd}, 
similarly to what was previously observed in $e^+e^-$ collisions \cite{L3}, an anticorrelation 
was reported.  This result required a study of different fitting functions, as had been suggested in 
the case of small systems, such as the $\tau$ model \cite{Htau}, which considers strong correlations between the space-time coordinates and the momentum components of the emitted particles, and was found to describe better the overall 
behavior of the correlation functions. 

A natural extension of those analyses is to investigate the GGLP/HBT correlations 
with respect  to different components of the pair relative momentum, which allows for 
exploring the source sizes in different directions.  Such analyses have also been
studies by other experiments at RHIC and LHC ~\cite{afs1,afs2,e735,star,phobos,phenix,alice}.  
Therefore, the HBT/GGLP correlation is measured in two-dimensions (2-D) as a function of the relative momenta along and transverse to the beam
direction,  $q_L$ and $q_T$, respectively. 
In three-dimensions (3-D) the Bose-Einstein correlations are studied  
in terms of $q_L$, $q_O$ and $q_S$, these last two obtained, respectively, by projecting the transverse component $\vec{q_T}$ in orthogonal directions, i.e.,  parallel to the average transverse momentum of the pair ($k_T$), and orthogonal to both $q_L$ and $q_O$. 

Using the same framework as in Ref. \cite{cms-hbt-1st,cms-hbt-2nd}, the  HBT/GGLP effect  is further scrutinised here for charged hadrons produced in minimum bias events in pp collisions at $\sqrt{s} = 2.76$ and 7~TeV  
with the data collected by CMS  
at the CERN LHC. New 1-D results at both these energies are discussed and then the analysis is extended to 2-D and 3-D cases.

\section{Bose-Einstein Correlation measurements} 
\label{sec:legacy-techni-compare}

\subsection{Event and Track Selections}
\label{subsec:evt-track-sel}

The data sets used in this analysis correspond to minimum bias samples obtained in pp collisions at 2.76 and 7~TeV recorded with the CMS detector at the LHC \cite{fsq-13-002-pas}. A detailed description of the CMS detector can be found in Ref. \cite{cms}. The minimum bias sample at $\sqrt{s}=$ 2.76~TeV was triggered on-line by requiring at least one track with $p_T > 0.4$~GeV to be found in the pixel tracker with $|\eta| < 2.4$ for a pp bunch crossing. Besides, in the offline analysis hadronic collisions were selected by requiring a coincidence of at least one Forward Hadronic (HF) calorimeter tower with more than 3~GeV of total energy in each of the HF detectors. 
In the case of pp collisions at 7~TeV events were selected by a trigger signal in each side of the Beam Scintillation Counter (BSC), coincident with a signal from either of the two detectors indicating the presence of at least one proton bunch crossing the interaction point (IP). 
Collision events were then selected offline by requiring a Beam Pickup for Timing for the eXperiments (BPTX) signal from both beams passing the IP.

The set of pp collision events at $\sqrt{s} = 2.76$~TeV used in this analysis comprises the data collected by CMS in 2013 at the CERN LHC  (3.4 million events). At $\sqrt{s} = 7$~TeV, a combined sample from pp collisions was considered, which uses data from three periods of the CMS data taking, i.e., commissioning run (23 million events), as well as from the runs 2010A (16 million events) and run 2010B (4 million events), where  the first is almost pileup free, 
while the later has a non-negligible fraction of events with multiple interactions. A filter was used for reducing the contamination in case of multiple vertices (the reconstructed vertex with the largest number of associated tracks is selected at 7~TeV, while in pp collisions at 2.76~TeV, an additional primary vertex might be identified as originating from a second pp collision by looking at its properties). 
To assess the related systematic uncertainty, an alternative event selection for reducing pileup contamination was also investigated by considering only single reconstructed vertex events.  

In the case of pp collisions at 2.76~TeV, three Monte Carlo samples were used. For obtaining the BEC results, minimum bias  events simulated with Pythia 6 Z2 Tune \cite{Pythia6.4,DT6} were employed, whereas Pythia D6T and Pythia Z2star \cite{Pythia6.4,DT6} were used for estimating the systematic uncertainties related to the choice of Monte Carlo tune. Each of them contained 2 million events. 
 For the analysis at 7~TeV about 33 million Monte Carlo events were simulated using Pythia 6 Z2 Tune. 

The track selection employed in both cases above follows  
the same criteria as in Ref. 
\cite{cms-hbt-1st,cms-hbt-2nd}, and are discussed in details in these references, as well as in \cite{fsq-13-002-pas}.

\subsection{The Bose-Einstein Correlation} 
\label{subsec:hbt-analysis}

The procedure adopted is the same as described in Refs.~\cite{cms-hbt-1st,cms-hbt-2nd}. Although no particle identification is considered in this analysis, the contamination from non-pions is not expected to be sizeable, as discussed in \cite{cms-hbt-1st,cms-hbt-2nd}, since pions are the dominant type of hadrons in the sample. 
For each event,  the signal containing the Bose-Einstein correlations is identified  
by pairing same charge tracks from the same event 
and distributing them in bins of the relative momentum of the pair, for instance, $q^\mu = k_1^\mu - k_2^\mu$, being $k^\mu_i$ the four-momenta of the individual particles in the pair. 
The background distribution or reference sample  
is formed similarly, by pairing charged particles from different events and within the same $\eta$ range (where  the full pseudorapidity interval is divided in three subranges $\Delta \eta$, corresponding to $-2.4 \le \eta \le -0.8$,  $-0.8 \le  \eta \le 0.8$, and  $0.8 \le \eta \le 2.4$), as in Ref.~\cite{cms-hbt-2nd}. This mixed event technique is referred to as ``same track density in $\Delta \eta$''. 
A single ratio is then formed, having the signal pair distribution as numerator and the reference sample as 
denominator, 
with the appropriate normalization, i.e., 
$R = \big( \frac{\cal{N}_{\textrm{ref}}}{\cal{N}_{\textrm{sig}}} \big)  \frac {(\textrm{d}N_\textrm{sig}/\textrm{d}Q_{\textrm{inv}} )}{(\textrm{d}N_{\textrm{ref}}/\textrm{d}Q_{\textrm{inv}})}. $  
The invariant relative momentum of the pair is defined as $Q_{\textrm{inv}}=\sqrt{-q^\mu q_\mu}=\sqrt{ -(k_1 - k_2)^2}$; $\cal{N}_{\textrm{sig}}$ is the integral of the signal pair distribution of all the events, 
whereas $\cal{N}_{\textrm{ref}}$ is the equivalent in the reference sample. 
A double ratio technique is then taken with the data and the Monte Carlo single ratios corresponding to the $Q_{\textrm{inv}}$ distributions, in terms of which the  
Bose-Einstein Correlation (BEC) effect 
is investigated \cite{cms-hbt-1st,cms-hbt-2nd}, 
${\cal{R}}(Q_{\textrm{inv}}) = \frac{R}{R_{\mbox{\tiny MC}}} =
\left(\frac{\mbox{d}N_\textrm{sig}/\mbox{d} Q_{\textrm{inv}}}{\mbox{d}N_{\mbox{\tiny ref}}/\mbox{d} Q_{\textrm{inv}}} \right)
\Big/
\left(\frac{\mbox{d}N_{\mbox{\tiny MC}}/\mbox{d} Q_{\textrm{inv}}}{\mbox{d}N_{\mbox{\tiny MC,\,ref}}/\mbox{d} Q_{\textrm{inv}}}\right), $
\noindent where $R_{\small {\textrm{MC}}}$ is the single ratio computed with the simulated
events generated without BEC. In each case, the reference samples for
data and simulation are obtained in the same way.  
This double-ratio procedure has the advantage of considerably reducing the sources of bias due to track 
inefficiency and other detector-related effects, as well as other Bose-Einstein correlations. 

The GGLP/HBT method reflects not only the quantum statistics of  the pair of identical particles, but is also sensitive 
to the underlying dynamics. In particular, in the case of charged hadrons, the correlation 
function may be distorted by strong, as well as by Coulomb interactions. For pions, the strong 
interactions can usually be neglected in femtoscopic measurements. 
As in Ref. \cite{cms-hbt-1st,cms-hbt-2nd}, 
the depletion (enhancement) in the correlation function caused by the Coulomb repulsion (attraction) of equal (opposite) charge pairs in the case of pions is corrected by weighting pair-wise with the inverse Gamow factor~\cite{noGamov}. This factor, in case of same charge and opposite charge, is given by 
$ G^{SS}_w(\eta_w) = \frac{2\pi\eta_w}{e^{2\pi\eta_w}-1}, ~~~ G^{OS}_w(\eta_w) = \frac{2\pi\eta_w}{1-e^{-2\pi\eta_w}}, $
\noindent with $\eta_w=\alpha_{em} m_{\pi}/Q_{\textrm{inv}}$, where $m_\pi$ the pion mass and $Q_{\textrm{inv}}$ the invariant relative momentum of the pair.

For performing the multidimensional analysis, the double ratios are investigated in terms of the projections of the relative momentum ${\bf q = k_1 - k_2}$ in two or three directions. In the 2-D case, the decomposition is made in $q_{L}$ ({\textit{longitudinal}} component, along the beam direction), and $q_{T}$ ({\textit{transverse}} component). In the 3-D case, additional projections are considered in the transverse plane, resulting in $q_O$ ({\textit{outwards}}), and $q_S$ ({\textit{sidewards}}), respectively along the average transverse momentum of the pair, $k_T = (k_{T_1} + k_{T_2})/2$, and orthogonal to it; $q_O$, $q_S$, and $q_L$ are mutually orthogonal. This decomposition of the relative momentum of the pair is also known as Bertsch-Pratt variables \cite{pratt,hamapad,bertsch}.  
The investigations  
are carried out in center-of-mass (CM), i.e., the LHC laboratory frame, 
as well as in the Local Co-Moving System (LCMS), characterized by the frame in which the longitudinal component of the pair average momentum ($ k_L = (k_{L_1} + k_{L_2})/2 = (k_{z_1} + k_{z_2})/2 $) is zero. Details about the 2-D and 3-D relative momentum projections, as well as the boost to the LCMS can be found in Ref.~\cite{mlisa-rev}. 

The parameterizations used to fit the correlation functions in one- (1-D), two- (2-D) and three-dimensions (3-D), respectively in terms of $Q_{\textrm{inv}}$, $(q_L, q_T) $ and $(q_S, q_L, q_O)$,  are listed below. Throughout this analysis $\hbar = c =1$ is adopted.

{\small
  \begin{equation}
    {\cal{R}}(Q_{\textrm{inv}} ) = C [ 1 + \lambda e^{ - (Q_{\textrm{inv}}  R_{\textrm{inv}})^a  } ] \; ( 1 + \delta \; Q_{\textrm{inv}}  ), 
    \label{eq:1d-levy}
    \end{equation}
}
{\small    
\begin{equation}
{\cal{R}}(q_L q_T) = C \Big\{ 1 + \lambda \exp{ \Big[ - \Big| q_T^2 R_T^2 + q_L^2 R_L^2 + 2 q_T q_L R_{LT}^2 \Big| ^{a/2} \Big] } \Big\} \times (1+ \alpha q_T + \beta q_L), 
\label{eq:2d-levy}\end{equation}
}
{\small
\begin{eqnarray}
\!\!\!\!\!\!\!\!{\cal{R}}(q_S, q_L, q_O) &=& C \Big\{ 1 + \lambda \exp{ \Big[ - \Big| q_S^2 R_S^2 + q_L^2 R_L^2 + q_O^2 R_O^2 + 2 q_O q_L R_{LO}^2 \Big|^{a/2} \Big] } \Big\} \times \nonumber \\
& &(1+ \alpha q_S + \beta q_L + \gamma q_O ).    
\label{eq:3d-levy}
\end{eqnarray}
}
In the above expressions, $\lambda$ is the intercept parameter (intensity of the correlation in the smallest bin of the pair relative momentum), $C$, $\delta, \alpha, \beta, \gamma$ are constants. The exponent  $a$ is the L\'evy index of stability satisfying the inequality $0 < a \le 2$. In all the above cases, if treated as a free parameter when fitting the double ratios, this exponent usually results into a number between the value characterizing the exponential ($a=1$) and the Gaussian ($a = 2$) functions. More details can be found in Ref. \cite{csorgohegyizajc}.  
For the sake of clarity, we denote the longitudinal component of the relative momentum in the CM frame as $q_L$, and in the LCMS,  as $q^*_L$. 
 In particular, in the case $a=1$  
the exponential term coincides with the Fourier transform of the source function $\rho(t,\vec{x})$, characterised by a Lorentzian distribution; the radius parameters $R_{\textrm{inv}}$, $(R_T, R_L)$, $(R_S, R_L, R_O)$, correspond to the lengths of homogeneity fitted to the correlation function in 1-D, 2-D and 3-D, respectively. The additional polinomial terms are introduced for accommodating  possible deviations of the baseline from unity at large values of  
 these variables (long-range correlations), as well as for allowing a better quality fit. In the 2-D and 3-D cases, $a=1$ leads to the so-called {\sl stretched exponential} function. 
In Eq. (\ref{eq:2d-levy}), $R_T = \bar{R}_T + \tau \beta_T \cos\phi$ and $R_L = \bar{R}_L + \tau \beta_L$  
where $\mathbf{\beta_T}=\frac{\mathbf{k_T}}{k^0}$ and $\beta_L=\frac{k_L}{k^0}$, originated in the mass-shell constraint ($q^\mu k_\mu =0 \rightarrow q^0=\frac{\mathbf{q}.\mathbf{k}}{k^0}$); $\phi$ is the angle between the directions of ${\bf q_T}$ and $\bf{\bf k_T}$ and $\tau$ is the source life-time. 
In Eq. (\ref{eq:3d-levy}) $R_S ^2 = \bar{R}_S^2$, $ R_L^2 = \bar{R}_L^2 + \tau^2 \beta_L^2$, $R_O^2 = \bar{R}_O^2 + \tau^2 \beta_T^2$ and  $R_{LO}^2 = \tau^2 \beta_L^2 \beta_T^2$.  Both in Eq. (\ref{eq:2d-levy}) and (\ref{eq:3d-levy}) a frame dependent cross-term, respectively proportional to $q_T q_L$ and $ q_O q_L$ is present.  However, when the analysis is performed in the LCMS, this cross-term does not contribute for sources symmetric along the longitudinal direction.  

\subsection{Systematic uncertainties} 
\label{subsec:systematics}

Various sources of systematical uncertainties were considered in this analysis, as listed in Table~\ref{tab:1d-syst-summary}, the first four being similar to what was discussed in 
Ref.~\cite{cms-hbt-1st,cms-hbt-2nd}. Two additional studies were also performed:  the effect of separating positive from negative charges in the single ratios (since $(+ +)$, as well as $(- -)$ charges are added in the signal; in the reference sample, besides these two combinations, the $(+ -)$ case is also added to the sample). The second study added is the effect of pileup events, investigated by comparing the results to the case where only single-vertex events are considered. The exponential function in Eq.(\ref{eq:1d-levy}) (L\'evy type with $a=1$) was adopted for this investigation. The total values of the systematic uncertainties are calculated by adding the individual contributions in quadrature. The systematic uncertainties estimated in 1-D and summarized in  Table~\ref{tab:1d-syst-summary} are extended  to both the 2-D and the 3-D cases.

\begin{table}[hbt]
\begin{center}
\footnotesize
\caption{ Spread with respect to the mean values at  $\sqrt{s}=2.76$~TeV and 7~TeV }
\begin{tabular}{l|c|c|c|c}
\hline
      \multicolumn{5}{c}{ Systematical Uncertainties} \\ \hline
  $\sqrt{s}$ 	& \multicolumn{2} {c|} {2.76~TeV} & \multicolumn{2} {c} {7~TeV} \\
\hline
Origin of Systematics  & $\lambda$ & $R_{\textrm{inv}}$ (fm)  & $\lambda$ & $R_{\textrm{inv}}$ (fm) \\
\hline
      Monte Carlo tune & 0.032   & 0.160  &   0.032   & 0.160  \\ 
      Reference Sample  & 0.009  &  0.047   & 0.051  &  0.188 \\ 
      Coulomb Corrections &  0.016   &  0.009   &  0.018 & 0.020 \\ 
      Track Cuts &  0.014  &    0.119   &  0.014 & 0.119  \\ 
      Charge Dependence &   0.006   &  0.012   &  0.007 & 0.006 \\ 
      Pileup filter  & 5.0 e-4  &   0.011   &  0.001 &  0.0025 \\ \hline
      Total  & 0.040  & 0.206   & 0.065  &  0.275 \\ \hline
\end{tabular}
 \label{tab:1d-syst-summary}
\end{center}
\end{table}

\section{Analysis Results}

\subsection{One-dimensional results} 
\label{subsec:legacy-results-1d}

\begin{figure}[h]
  \begin{center}
\includegraphics[width=0.38\textwidth]{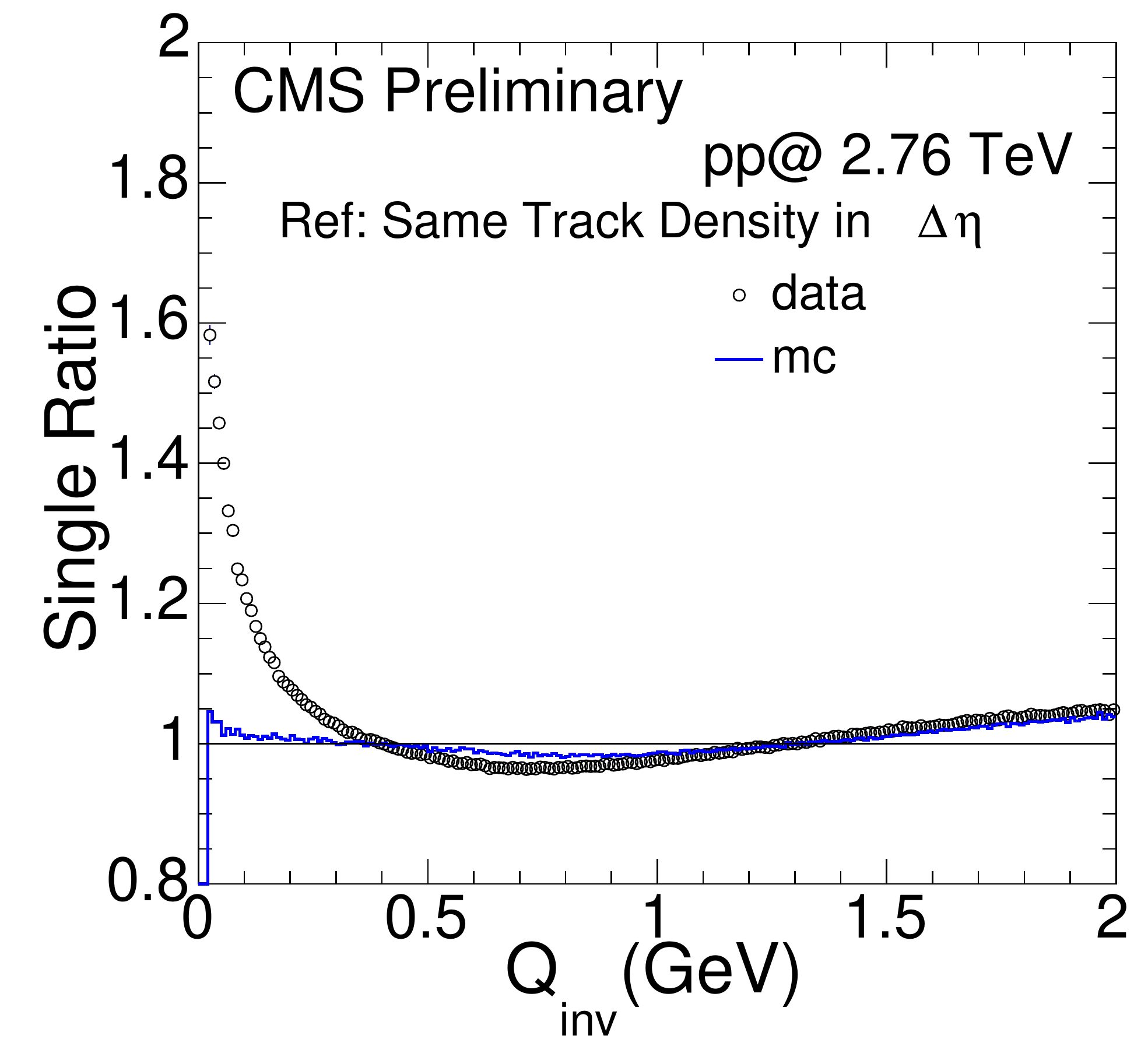}
\includegraphics[width=0.38\textwidth]{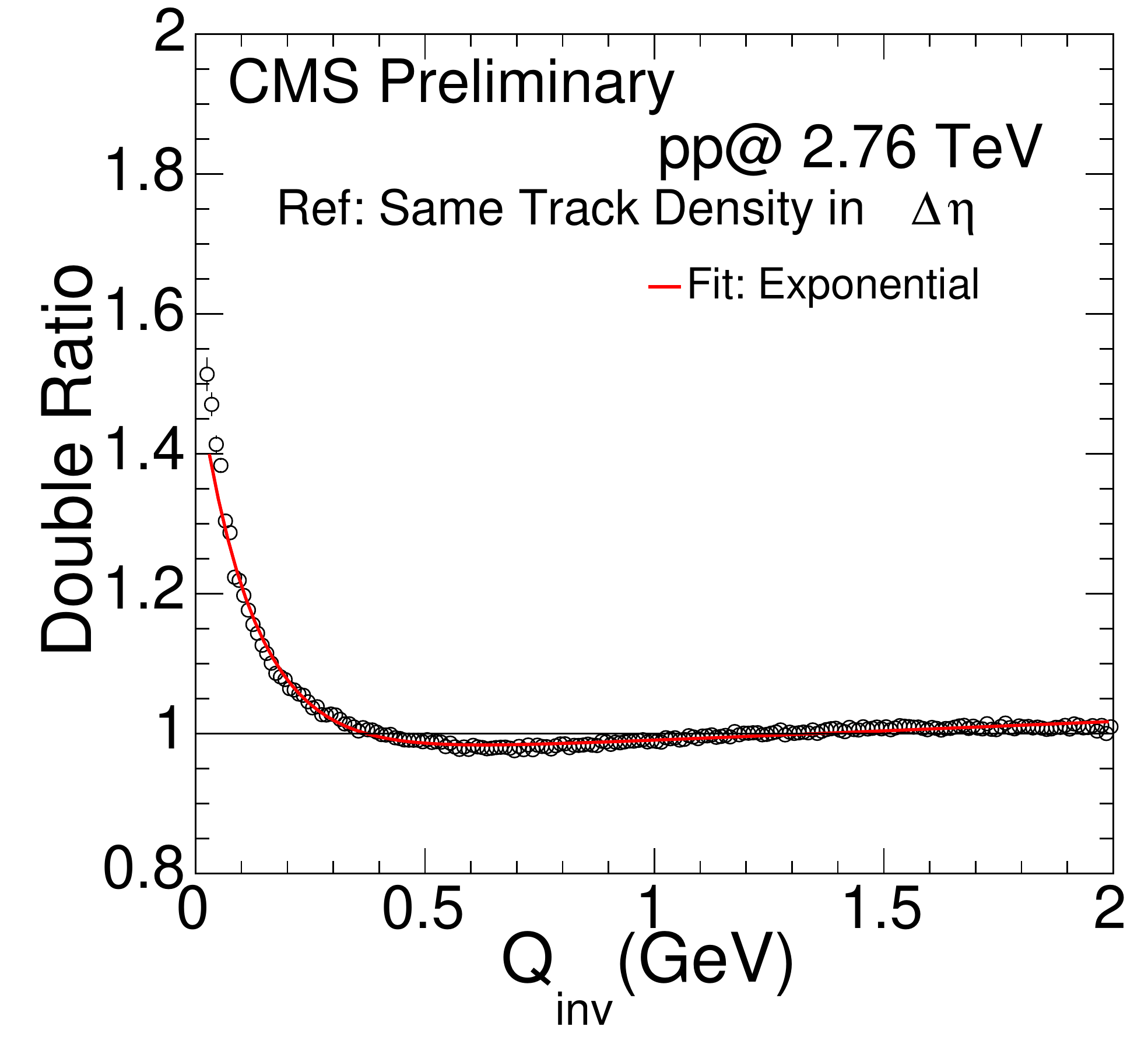}
  \caption{ 1-D single ratios as a function of $Q_{\textrm{inv}}$ for data and Monte Carlo (Pythia 6-Z2 tune) 
  from to  pp collisions at 2.76 TeV are shown (left), as well as the corresponding double ratio superimposed by the exponential fit (right). }
   \label{fig:1-d-singleratios}
\smallskip
    \includegraphics[width=0.44\textwidth]{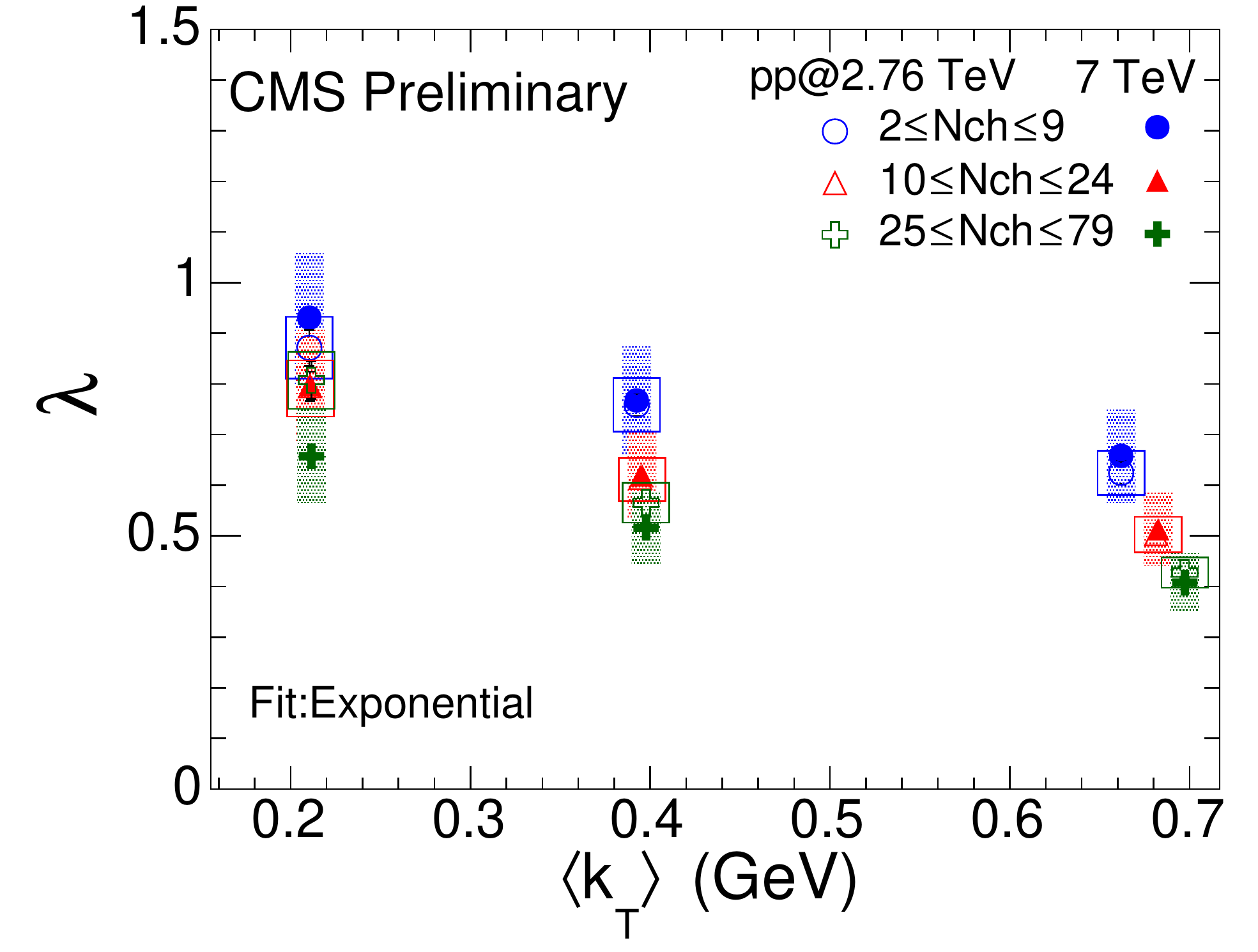}
    \includegraphics[width=0.44\textwidth]{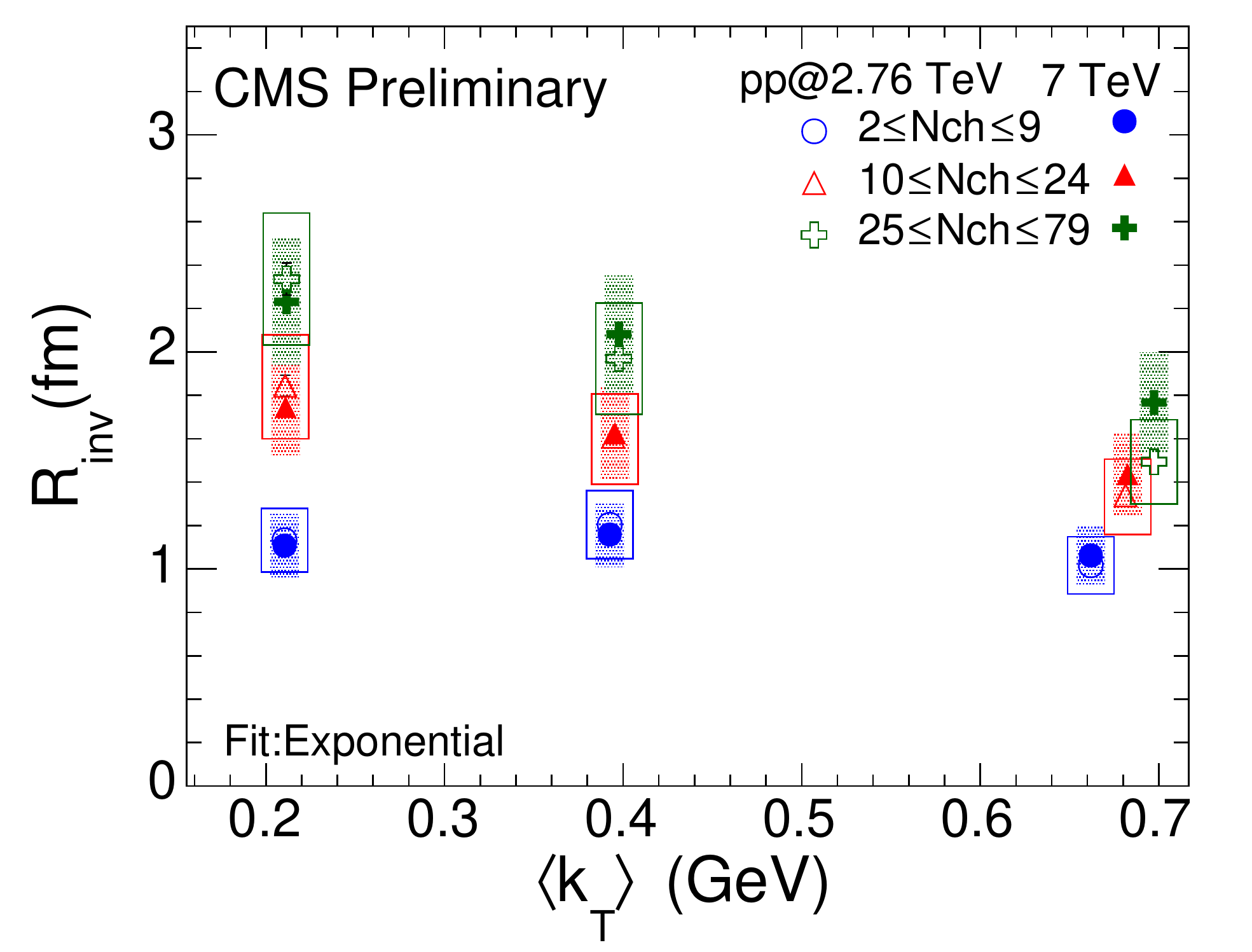}
    \caption{The fit parameters $\lambda$ (left) and $R_{\textrm{inv}}$ (right)
    from exponential fits to the double ratios are shown in different $N_{ch}$ and $k_T$ bins, from pp 
    collisions at 2.76 and 7 TeV (full sample). 
    The statistical uncertainties are indicated by error bars (in some     cases, smaller than the marker's size),  the systematic ones by empty (2.76 TeV data) or shaded boxes (7 TeV     data). }
        \label{fig:fits-to-dr-nch-kt-276-7tev}
\smallskip
    \includegraphics[width=0.41\textwidth]{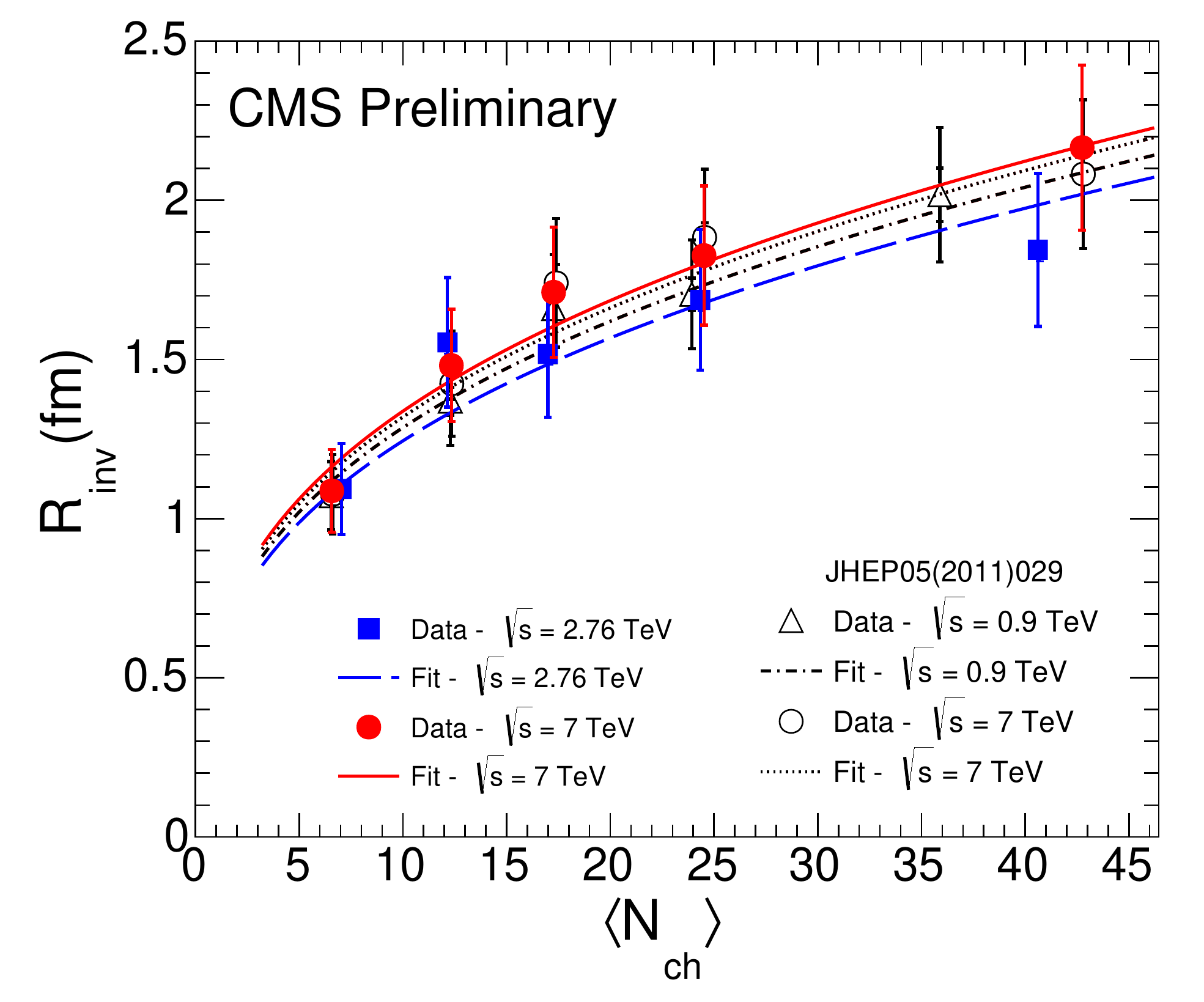}
     \includegraphics[width=0.41\textwidth]{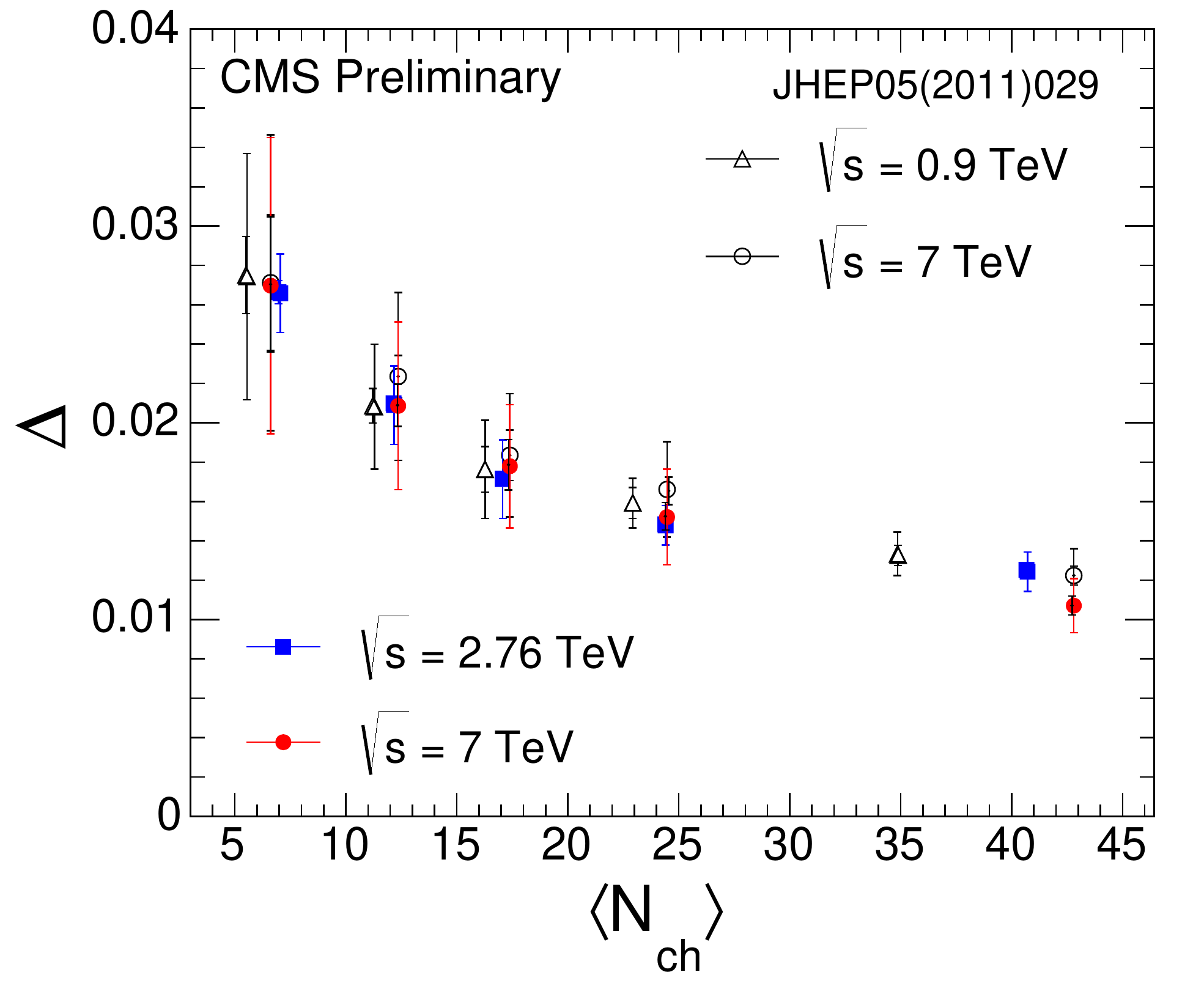}
    \caption{ Comparative plots with results in Ref.~\cite{cms-hbt-2nd}. 
    Left:  $R_{\textrm{inv}}$ versus $<N_{ch}>$ (acceptance and efficiency corrected), for pp 
    collisions at 2.76 and 7 TeV (fit curves are proportional to $N_{ch}^{1/3}$). The inner error bars represent 
    statistical uncertainties and the outer ones, statistical and systematic uncertainties added in quadrature. Right: The 
    anticorrelation's depth, $\Delta$, versus $<N_{ch}>$. } 
    \label{fig:dr-radii-nch-dipdepth-276-7}
  \end{center}
\end{figure}

The current analysis extends the previous one for pp at $\sqrt{s} = 7$ TeV reported in Ref.~\cite{cms-hbt-2nd} to full data sample, as well as at $\sqrt{s} = 2.76$ TeV with full available statistics. The corresponding results for single and double ratios are shown in Fig.~\ref{fig:1-d-singleratios}. The fits to the double ratios were produced with the exponential function in Eq. (\ref{eq:1d-levy}), with $a=1$. 

Figure~\ref{fig:fits-to-dr-nch-kt-276-7tev} shows results for the intercept parameter $\lambda$
and the invariant radius $R_{\textrm{inv}}$  
from pp collisions at 2.76 and 7~TeV (full statistics), 
in terms of $N_{ch}$ and $k_T$. It can be seen that the results corresponding to 
the two energies are very similar 
for the different ($N_{ch}$, $k_T$) combinations. 
The intercept $\lambda$ 
decreases with increasing $k_T$ and $N_{ch}$, whereas 
$R_{\textrm{inv}}$ steadily increases with multiplicity and seem to decrease with $k_T$, at least for the two largest $N_{ch}$ bins, showing that the lengths of homogeneity accessible to interferometric measurements decrease with the average pair momentum, as has been previously observed in several different collision systems and energy ranges \cite{cms-hbt-1st,cms-hbt-2nd,e735,star,phobos,phenix,alice}.This behavior is expected in the case of emitting sources originated from expanding systems.

In Fig.~\ref{fig:dr-radii-nch-dipdepth-276-7} (left), $R_{\textrm{inv}}$ is investigated as a function of 
the charged particle multiplicity, $<N_{ch}>$ (efficiency and acceptance corrected), where the curves are fits proportional to $N_{ch}^{1/3}$. 
The results for pp collisions at 2.76 TeV and at 7 TeV are consistent with former studies reported in Ref.~\cite{cms-hbt-2nd} at 0.9 and 7 TeV, showing a similar increase with $N_{ch}^{1/3}$. Such results also suggest an approximate scaling property of the lengths of homogeneity with increasing collision center-of-mass energy.

As discussed in \cite{cms-hbt-2nd}, an anticorrelation (value below unity) was observed in the double ratios for values of $Q_{\textrm{inv}}$ away from the Bose-Einstein peak, whose depth  was shown to decrease with increasing $N_{ch}$, for integrated values in $k_T$. This dip structure is also observed in the present analysis and its depth is further investigated. More details and discussion on the results are presented in Ref.~\cite{fsq-13-002-pas}. 
An additional function, $ {\cal{R}}(q) = C \left\{ 1+ \lambda [{\textrm{cos}}\left[ (q r_0)^2+\textrm{tan}(\alpha\pi/4)( q r_\alpha)^{\alpha}\right] e^{-(qr_\alpha)^\alpha}] \right\} \cdot(1+\delta q)$, was used to fit the data points, which  better describes such anticorrelation, as discussed in \cite{cms-hbt-2nd, fsq-13-002-pas}. It is based on the so-called $\tau$ Model~\cite{Htau}, which parameterizes  the time evolution of the source by means of a one-sided asymmetric 
L\'evy distribution. 
The {\sl dip}'s depth  \cite{cms-hbt-2nd} is estimated by the difference of the base-line function, $C(1+\delta q)$, and the remaining fit function based on the $\tau$ model at its minimum, leading to the results shown in Fig.~\ref{fig:dr-radii-nch-dipdepth-276-7} (right), where the results for pp collisions at both 0.9 and 7 TeV from Ref.~\cite{cms-hbt-2nd} and the new ones for 2.76 TeV and the full sample at 7 TeV are shown together (see \cite{fsq-13-002-pas} for details). 

\clearpage

\subsection{Two-Dimensional Results}
\label{subsec:doubratio_results-2d}

The BEC analysis is extended to the 2-D case in terms of the components $q_L, q_T$ of the pair relative momentum, with the data samples from pp collisions at $\sqrt{s}=2.76$ TeV and 7 TeV. It is performed both in the CM frame and in the LCMS, in which 
the cross-term depending on $q_Tq_L$ in 2-D, or $q_L q_O$, in 3-D, does not contribute, in case of longitudinally symmetric systems.

\begin{figure}[hbt]
  \begin{center}
    \includegraphics[width=0.44\textwidth]{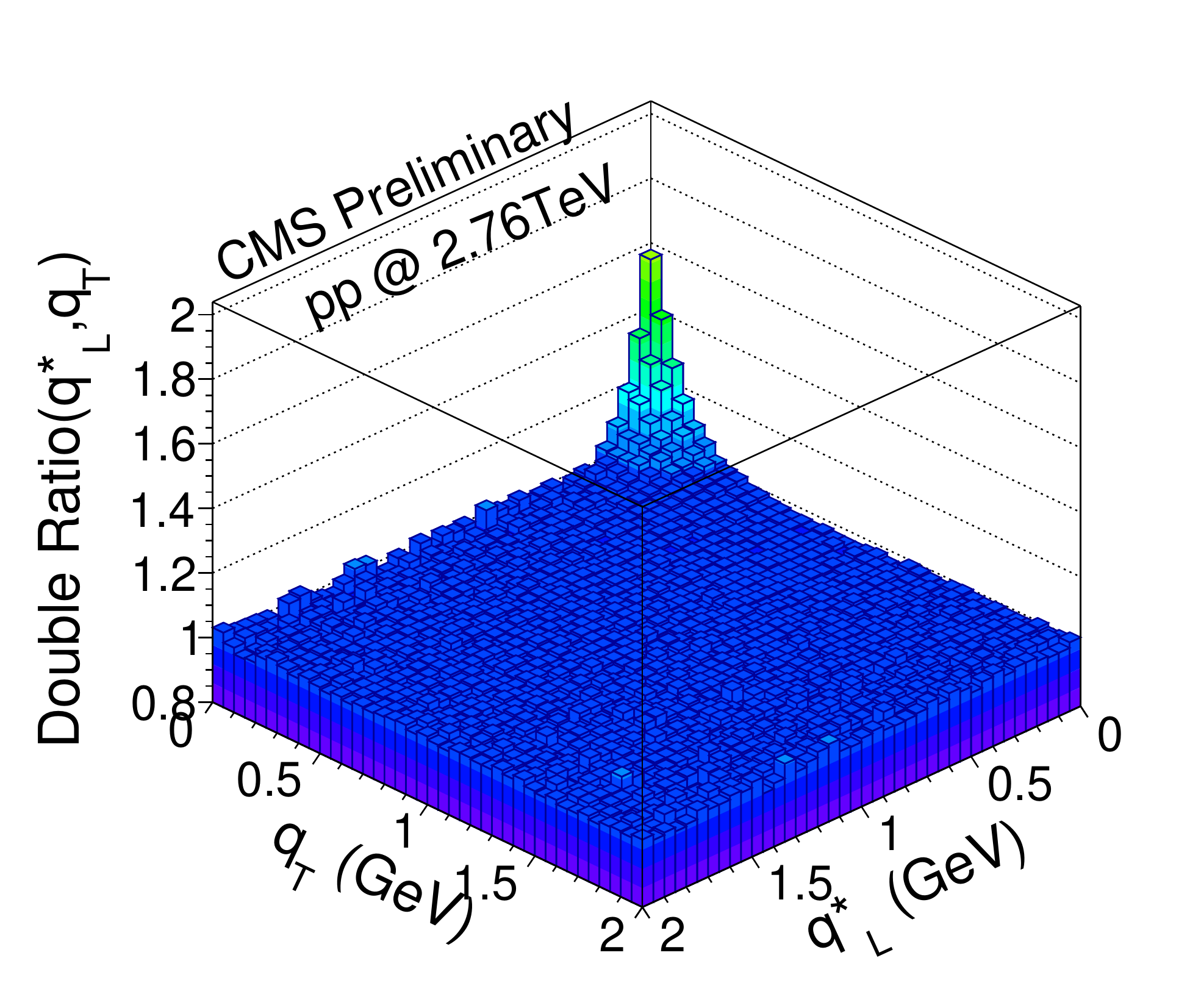}
       \includegraphics[width=0.51\textwidth]{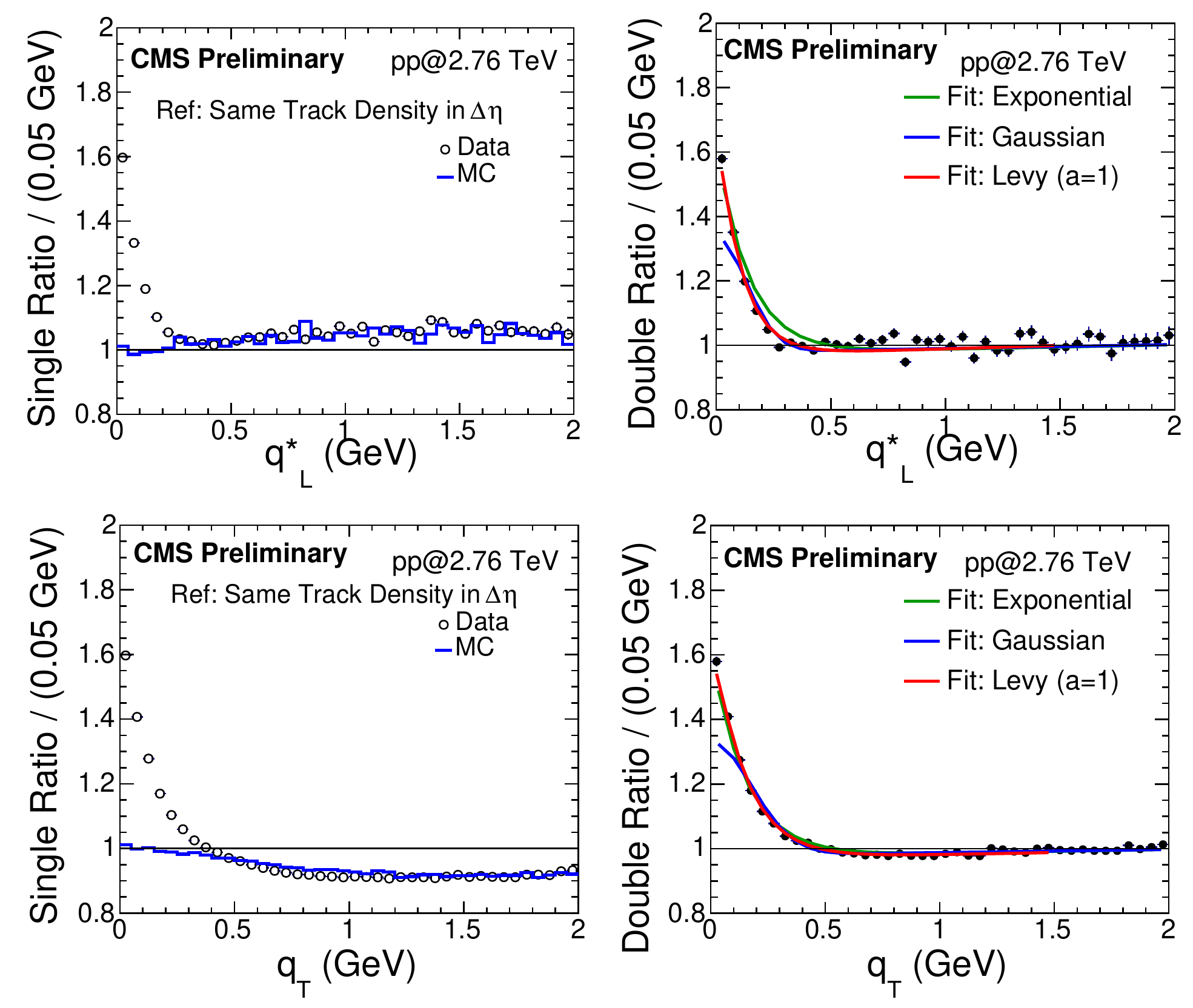} 
    \includegraphics[width=0.44\textwidth]{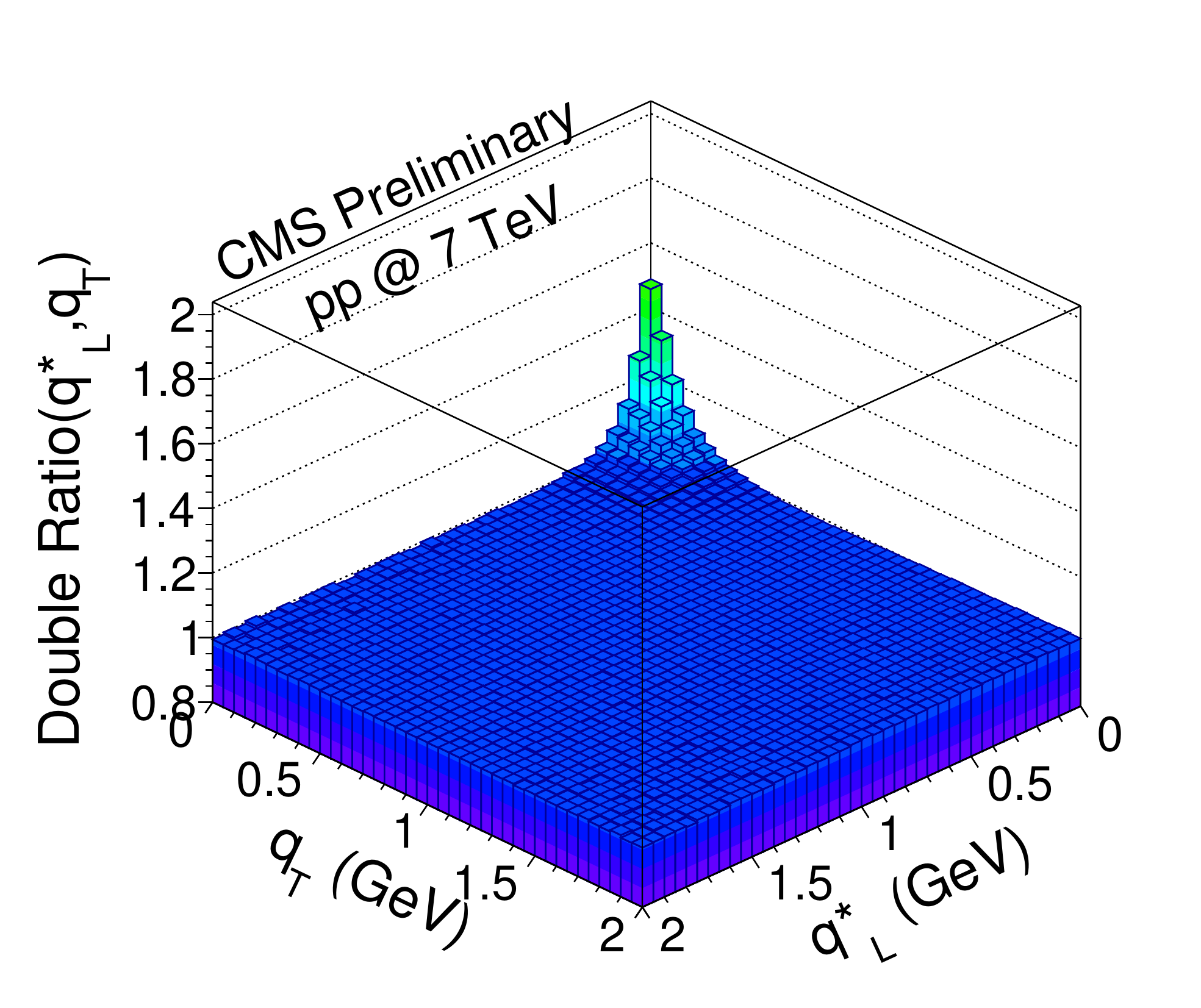}
       \includegraphics[width=0.51\textwidth]{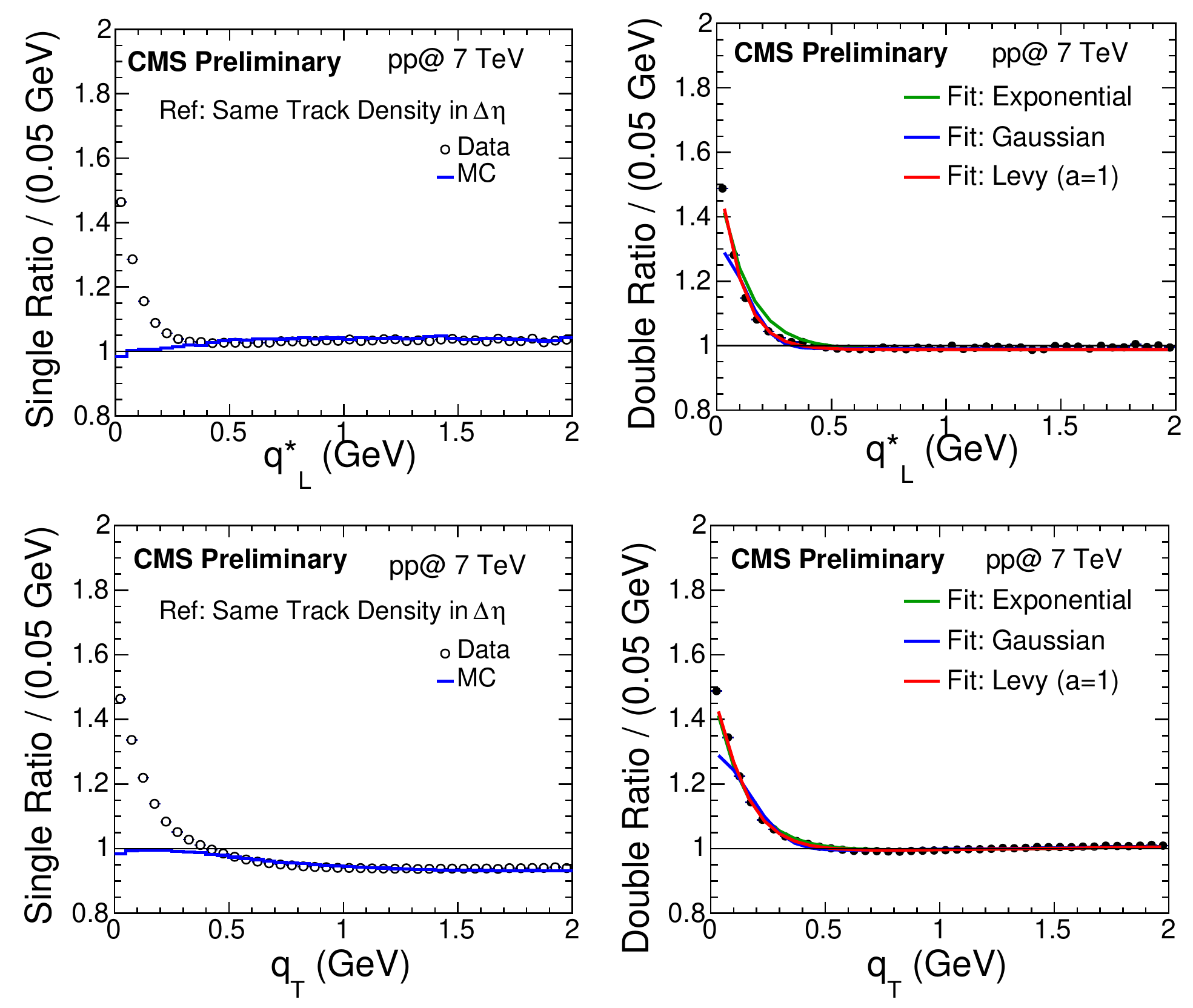}
    \caption{ The left panel shows double ratios as a function of ($q^*_L, q_T$), with data from pp collisions at 
    2.76 TeV (top) and 7 TeV (bottom) in the LCMS corresponding to results integrated in all $N_{ch}$ and $k_T$ bins. 
    The right panel shows the corresponding 1-D projections of the single and double ratios in terms of $q^*_L$ 
    (for $|q_T| < 0.05$ GeV) and $q_T$ (for $|q^*_L| < 0.05$ GeV). Gaussian, exponential and L\'evy (with $a=1$) 
    fit functions are shown superimposed to the data points.  }
    \label{fig:2d_doubleratios-project-276-7tev-lcms}
  \end{center}
\end{figure}

As an illustration, the double ratios in the LCMS are shown in Fig.~\ref{fig:2d_doubleratios-project-276-7tev-lcms}, as a 2-D plot (left panel) in terms of $(q^*_L,q_T)$, and the corresponding 1-D projections (right panel) 
for pp collisions at 2.76 TeV (top), and for 7 TeV (bottom). The 1-D projections, when plotted in terms of  $q^*_L$, considers only the first bin in $q_T$ (i.e., $q_T<0.05$ GeV), and vice-versa, with the data superimposed by the Gaussian, exponential and L\'evy (with $a=1$) fit functions. 
 
\begin{figure}[hbt]
  \begin{center}
  \includegraphics[width=0.327\textwidth]{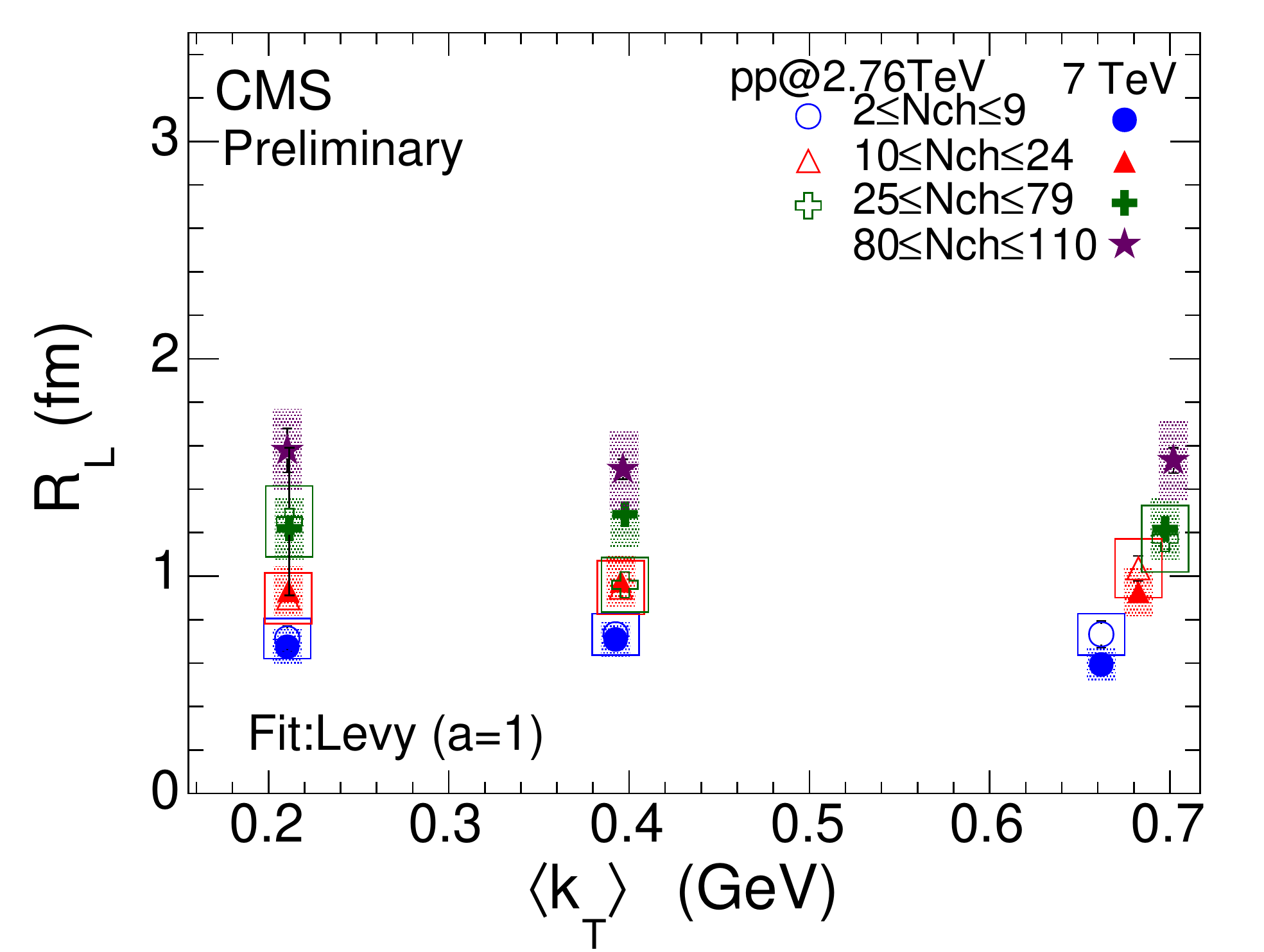}
  \includegraphics[width=0.327\textwidth]{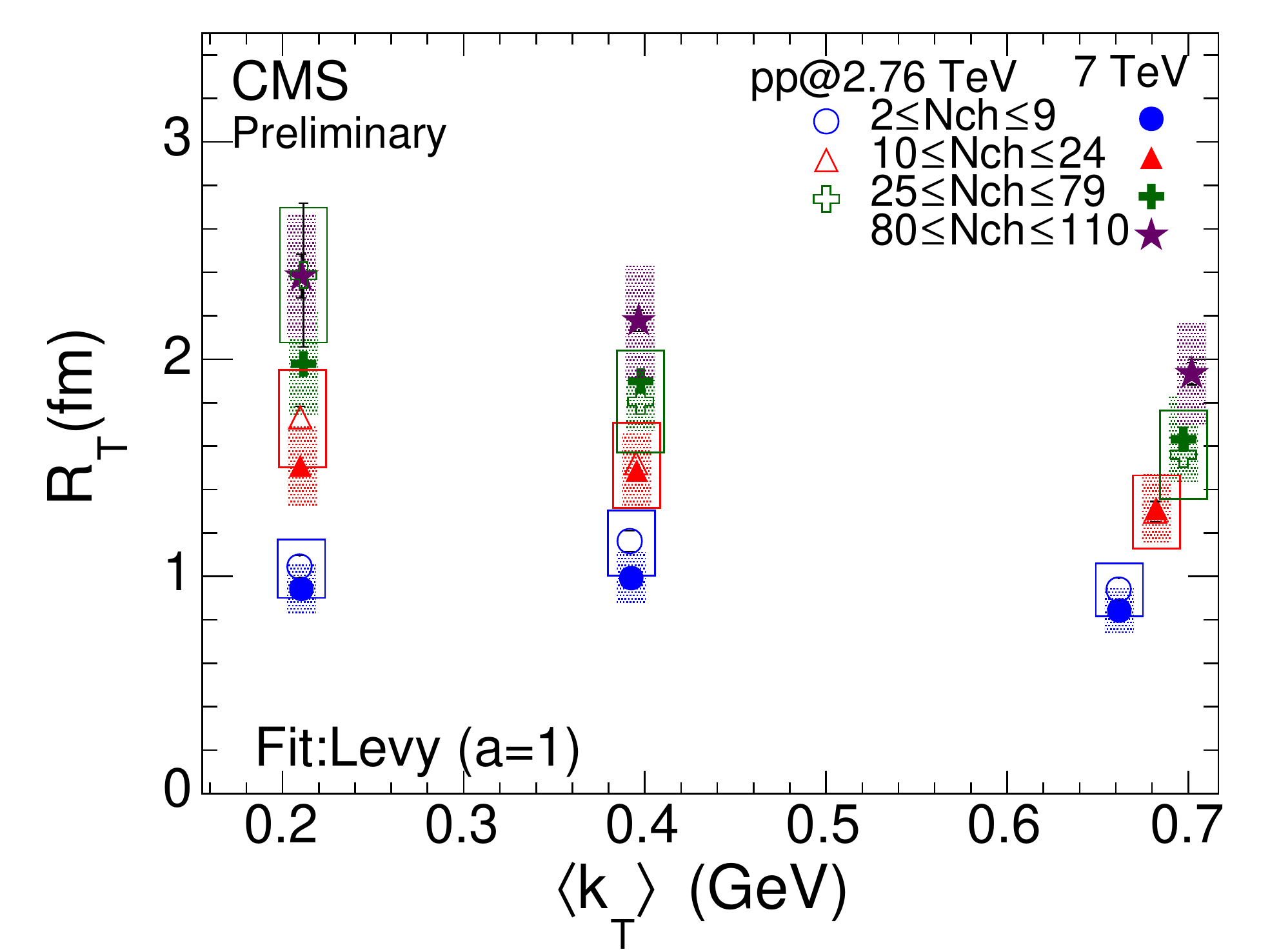}
  \includegraphics[width=0.327\textwidth]{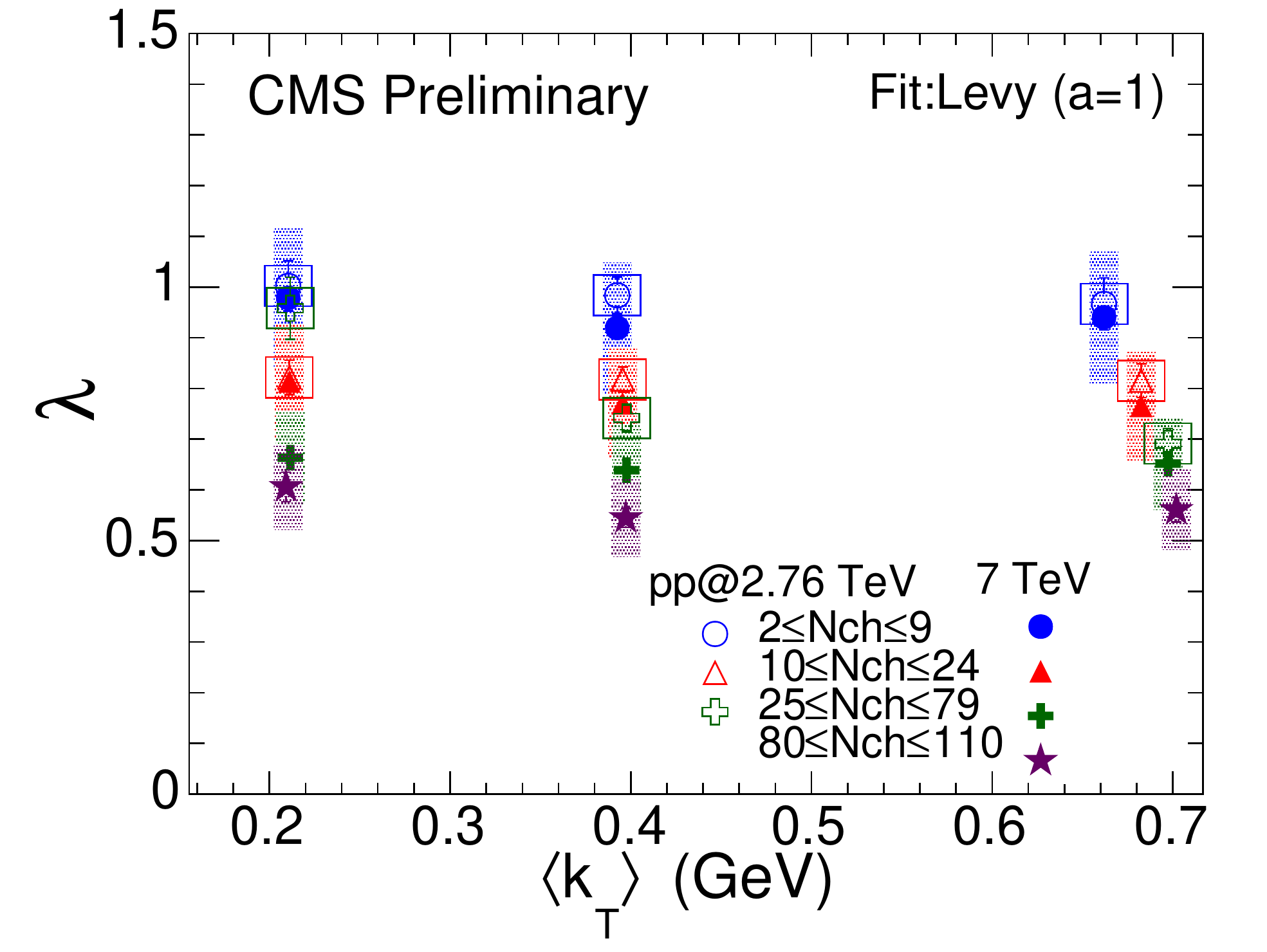} 
  \includegraphics[width=0.327\textwidth]{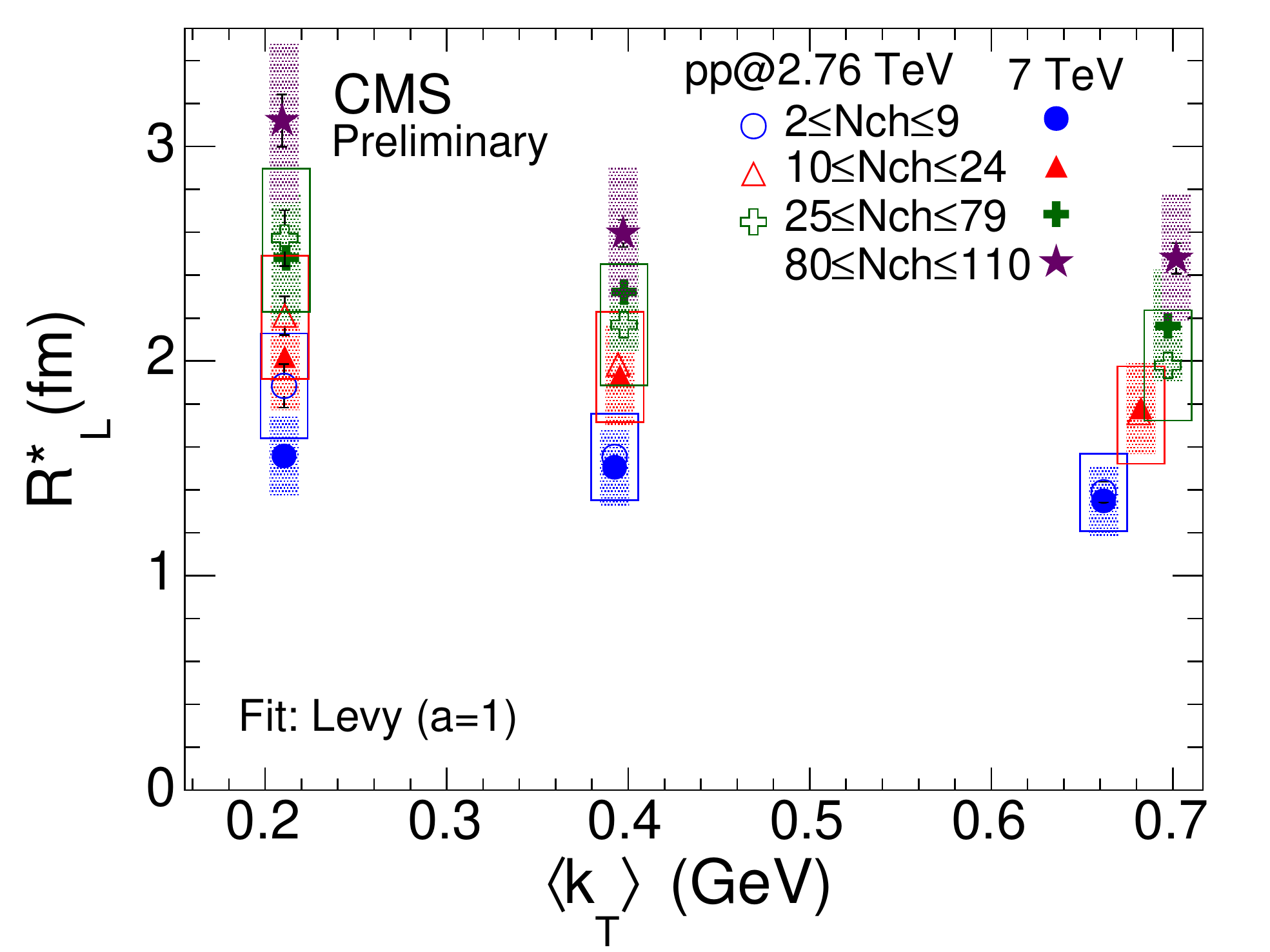} 
  \includegraphics[width=0.327\textwidth]{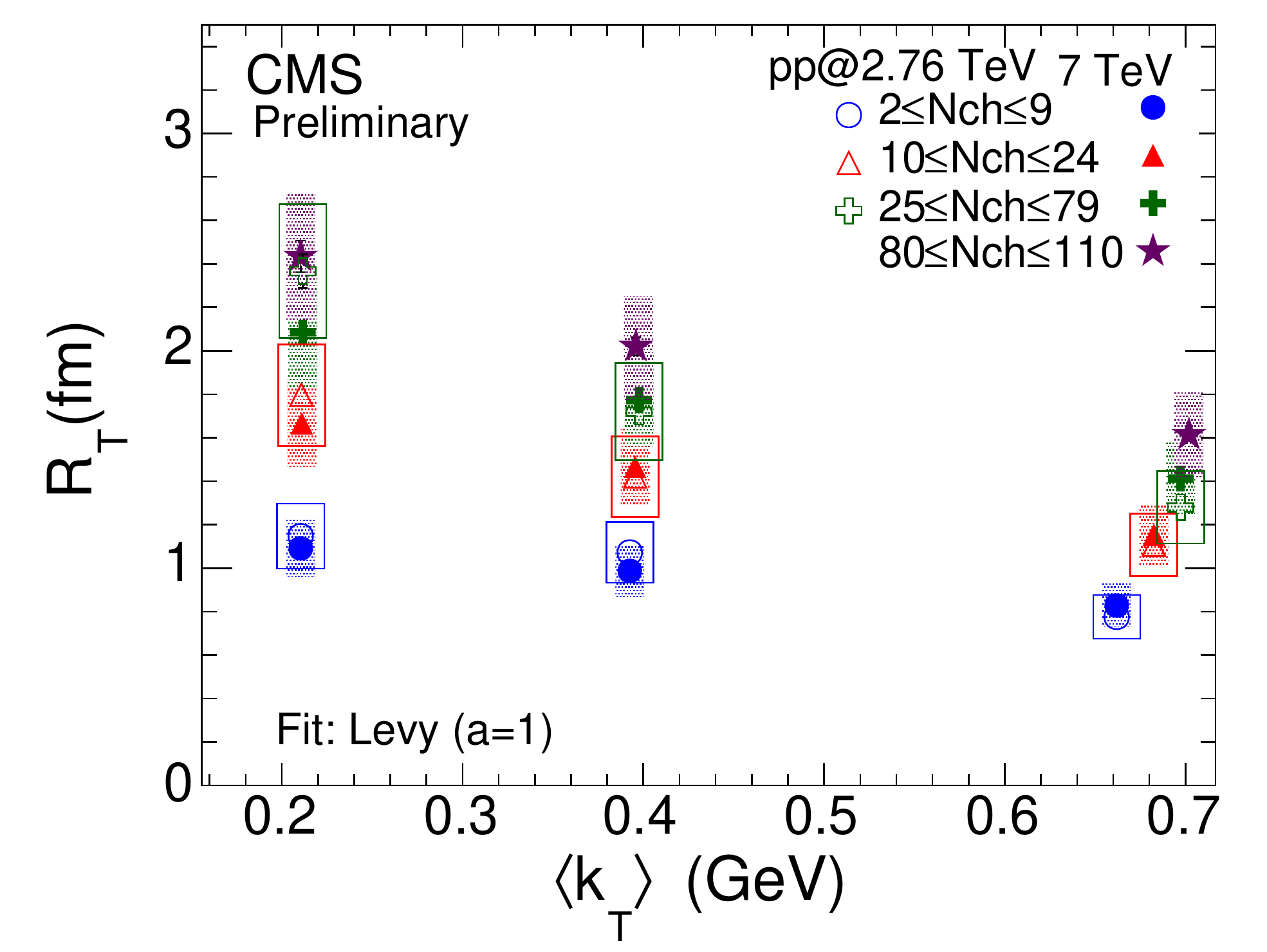}
  \includegraphics[width=0.327\textwidth]{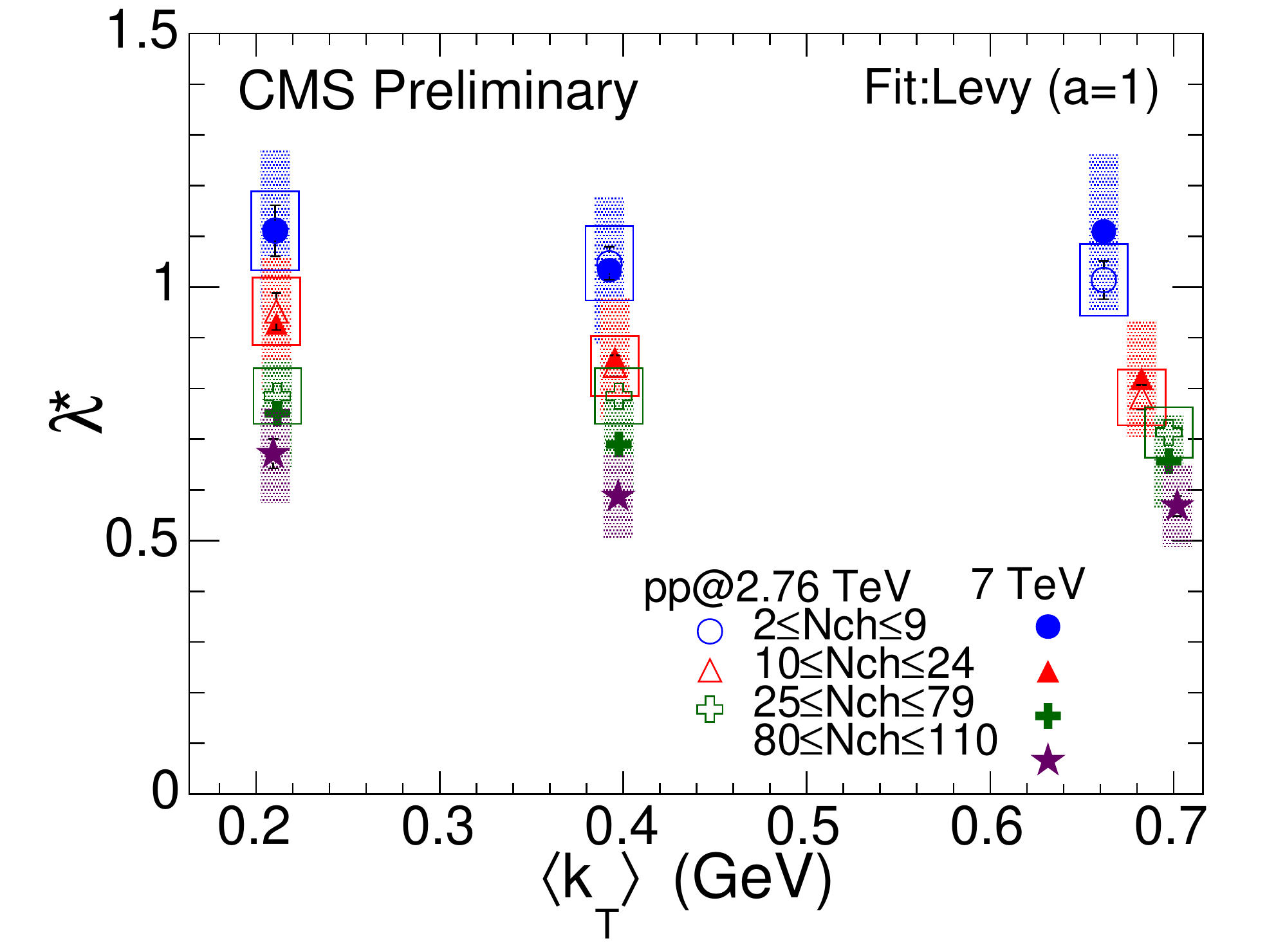}
   \caption{ The fit parameters obtained with the {\sl stretched exponential}  (L\'evy-type with $a=1$) function are shown for pp collisions at 2.76 and 7 TeV in the CM frame (top) and in the LCMS (bottom), as a function of $\langle k_T \rangle$ and for different 
   $N_{ch}$ bins. 
   The statistical uncertainties are indicated by error bars (in some 
   cases, smaller than the marker's size), whereas the systematic ones are indicated by empty (at 2.76 TeV) or shaded boxes (at 7 TeV). }
   \label{fig:2d_radii_rl_rt-lambda-276_7tev-cm-lcms}
   \end{center}
\end{figure}

Analogously to the studies performed in 1-D 
the double ratios in 2-D were also investigated in terms of $(q_L, q_T)$, in three 
intervals of the pair average momentum, $k_T$, and in different $N_{ch}$ bins. 
The results from the {\sl stretched exponential} fit (L\'evy-type with $a=1$) \cite{csorgohegyizajc} to the double ratios, performed both in the CM frame (top) and in the LCMS (bottom), are compiled in Fig.~\ref{fig:2d_radii_rl_rt-lambda-276_7tev-cm-lcms}.  The behaviour of the directional lengths of homogeneity is very similar in both frames, 
with $R_L$ ($R^*_L$) and $R_T$ increasing with charged multiplicity, $N_{ch}$,  
and decreasing with the average transverse momentum, $k_T$, at least in the larger multiplicity bins, a behaviour similar to that observed in the 1-D case and expected for  expanding sources. 
Another interesting feature of the data than can be observed in Fig.~\ref{fig:2d_radii_rl_rt-lambda-276_7tev-cm-lcms} is that $R^*_L$ (LCMS) $>R_L$ (CM) for the same bins of $N_{ch}$ and $k_T$, suggesting an effect of Lorentz boost contraction in the longitudinal length of homogeneity in the CM frame.  
Regarding 
$\lambda$, as shown in Fig.~\ref{fig:2d_radii_rl_rt-lambda-276_7tev-cm-lcms},  
no significant sensitivity of the intercept is seen as a function of 
$k_T$.  However, within each $k_T$ range, $\lambda$ slowly decreases with increasing track multiplicity in an similar way in both frames.

\begin{table}[hbt]
\begin{center}
\footnotesize
\caption{ 2-D fit parameters for in the LCMS }
\begin{tabular}{l|c|c}
\hline
\multicolumn{1} {c|} {$\sqrt{s}$ }   	& \multicolumn{1} {c|} {2.76 TeV} & \multicolumn{1} {c} {7 TeV} \\  \hline
\multicolumn{1} {c|} {$\lambda$}  &   $ 0.830 \pm 0.010 $  (stat.)  $ \pm~0.040 $ (syst.)     &    $0.700 \pm 0.002$  (stat.) $ \pm~0.065 $ (syst.)    \\  \hline
\multicolumn{1} {c|} {$R_T$ (fm)}     &  $ 1.498 \pm 0.013 $ (stat.)  $ \pm~ 0.206$ (syst.)   &   $1.640 \pm 0.003$ (stat.)  $ \pm~0.206$ (syst.)    \\ \hline
\multicolumn{1} {c|} {$R^*_L$ (fm)}     &  $1.993 \pm 0.022$ (stat.)  $ \pm~0.206$ (syst.)   &   $ 2.173 \pm 0.005 $ (stat.)  $ \pm~0.275$ (syst.)   \\ \hline
\end{tabular}
 \label{tab:radii-276-7tev}
\end{center}
\end{table}

Table~\ref{tab:radii-276-7tev} collects the values of the radius, $R^*_L, R_T$, and of the intercept, $\lambda$, fit parameters, integrated both in $N_{ch}$ and $k_T$,  
and obtained with the {\sl stretched} exponential fit (i.e., L\'evy-type with $a=1$). 
From Table~\ref{tab:radii-276-7tev} it can be seen that, in the LCMS, the rest frame of the pair along the longitudinal direction, $R^*_L \approx 4 R_T/3$, suggesting that the source is longitudinally elongated, at both energies. 

\begin{figure}[hbt]
  \begin{center}
   \includegraphics[width=0.48\textwidth]{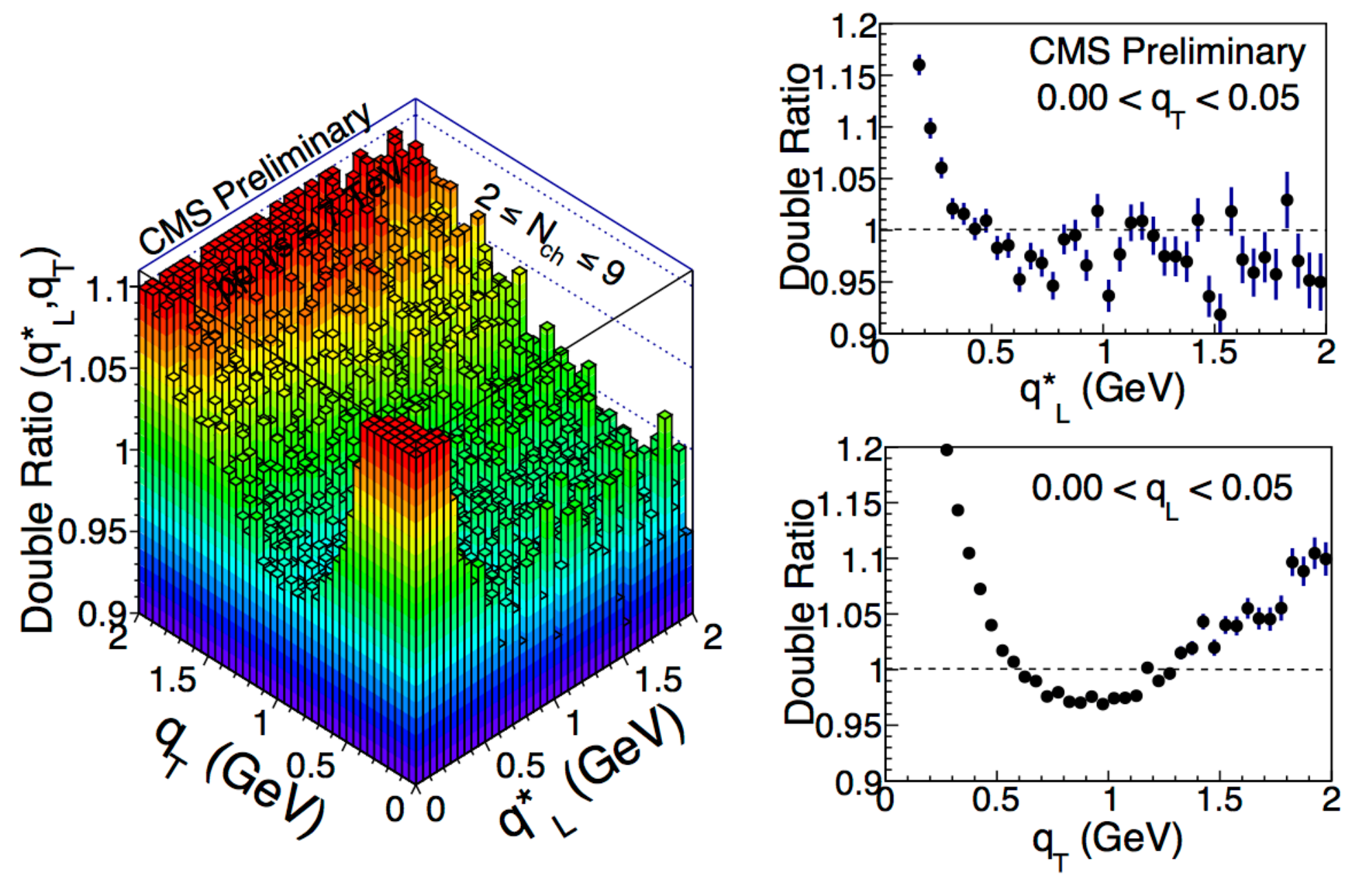}
   \includegraphics[width=0.48\textwidth]{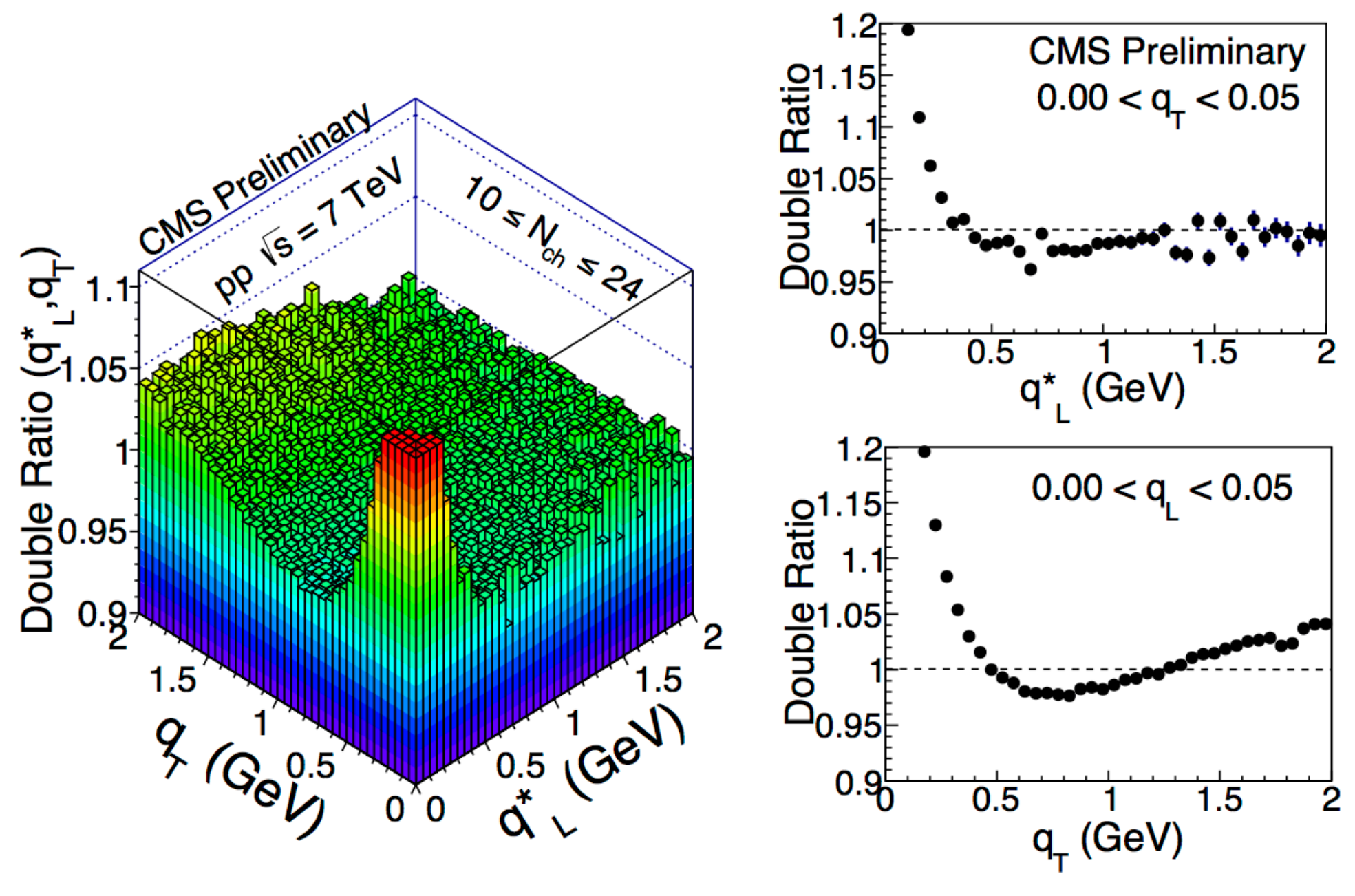} 
   \includegraphics[width=0.48\textwidth]{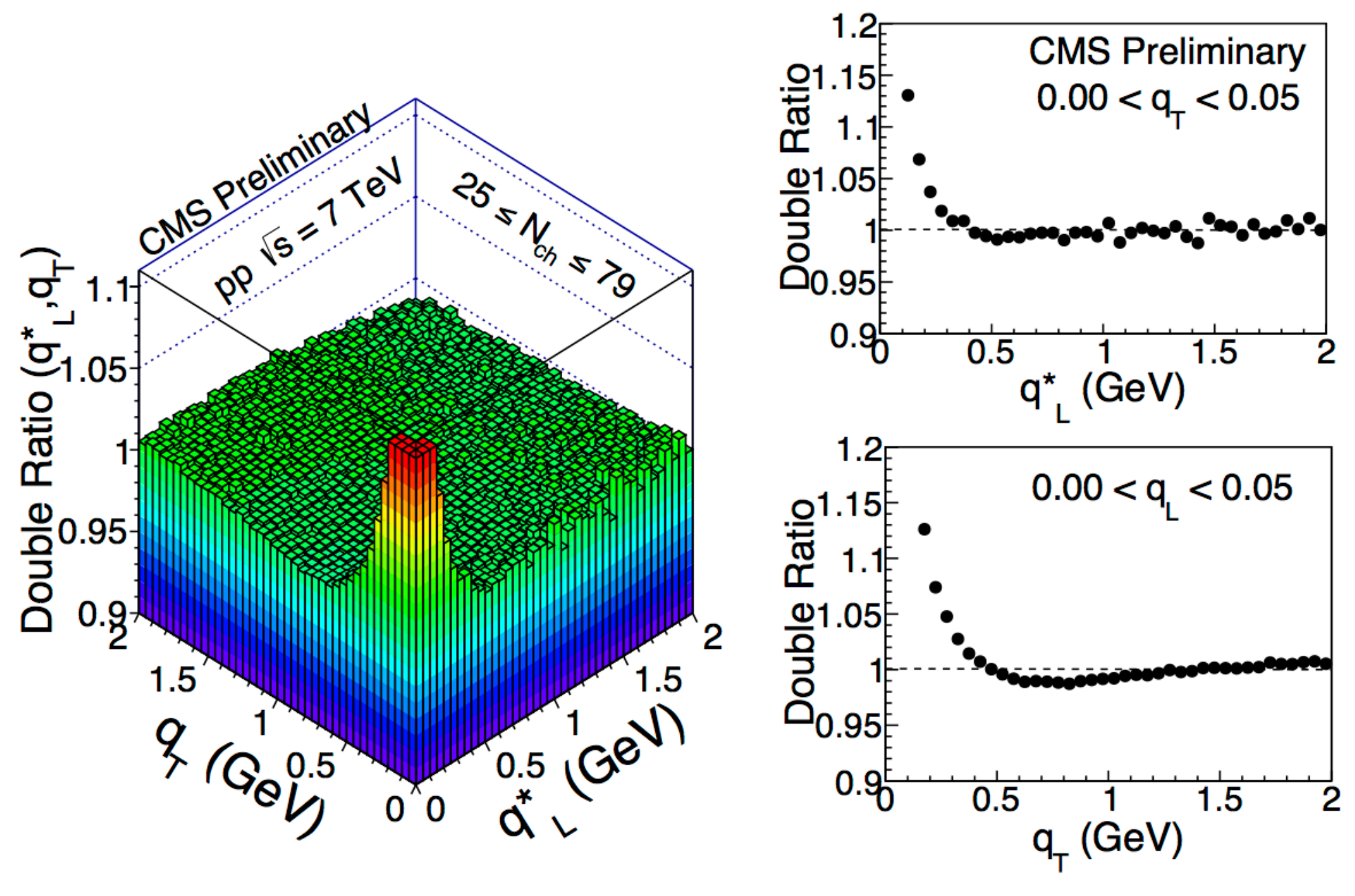}
   \includegraphics[width=0.48\textwidth]{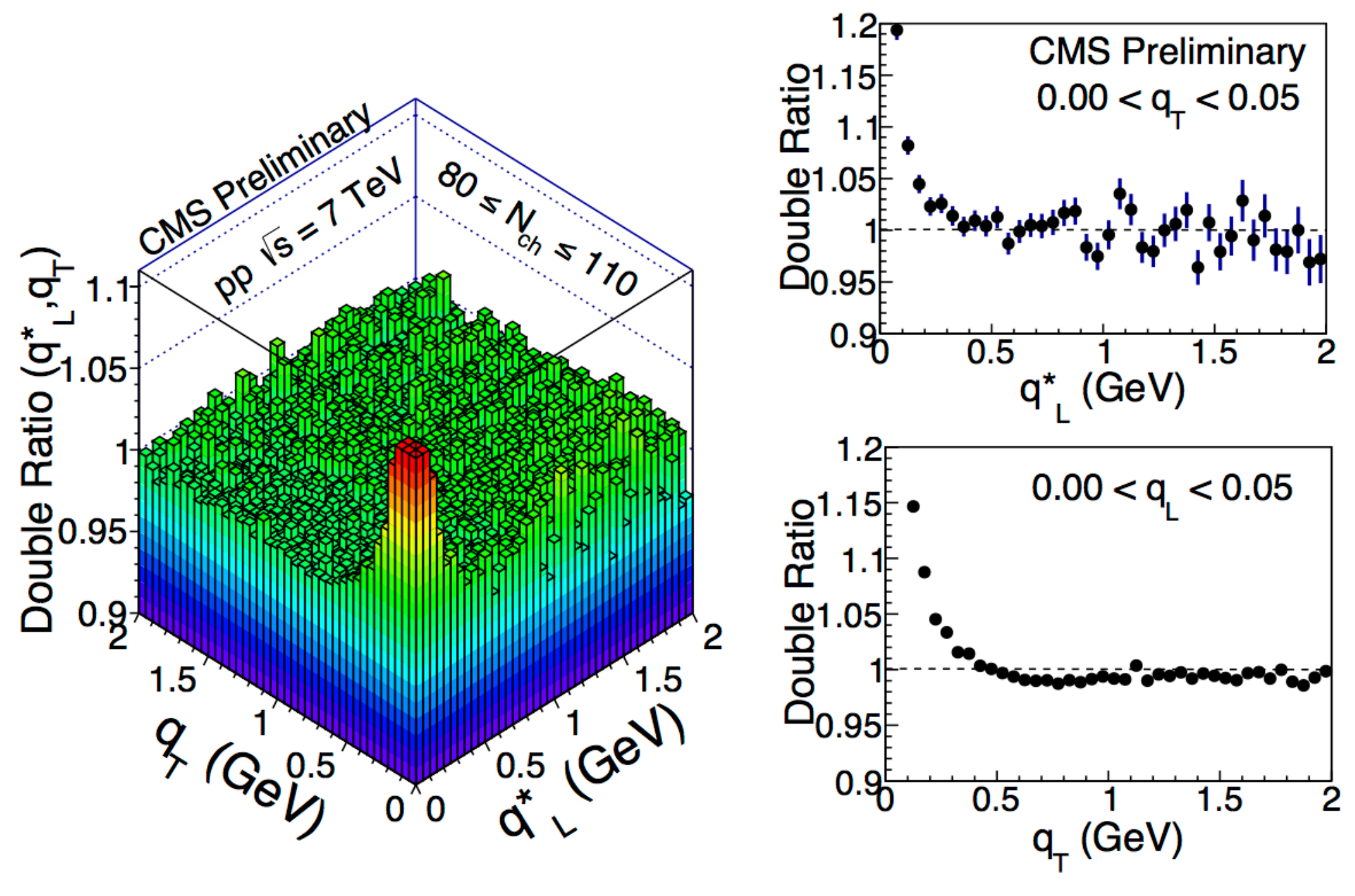}
\caption{ Results obtained in the LCMS for the 2-D double ratios with zoomed axes, with the BEC peak cut above 1.2, as a function of ($q^*_L, q_T$) for four charged multiplicity bins, $N_{ch}$, increasing from top left to bottom right.  The 1-D projections in $q^*_L$ (for $|q_T| < 0.05$ GeV) and $ q_T$ (for $|q^*_L| < 0.05$ GeV) are shown side-by-side to the corresponding 2-D double ratios.  }
  \label{fig:2d-dip-structure-lcms} 
   \end{center}
\end{figure}

In Fig.~\ref{fig:2d-dip-structure-lcms} the 2-D results for the double ratios versus ($q^*_L, q_T$) in the LCMS are zoomed along the correlation function axis, which cuts the BEC peak above 1.2. Figure~\ref{fig:2d-dip-structure-lcms} also shows the 1-D projections in terms of each of these variables, having the complementary one within the first bin, i.e, $|q_i| < 0.05$ GeV. The results are shown in four bins of charged particle multiplicity, $N_{ch}$, which increases from the top left panel to the bottom right one. 

\subsection{Three-Dimensional Results}
\label{subsec:doubratio_results-3d}

\begin{figure}[h]
  \begin{center}
    \includegraphics[width=\textwidth]{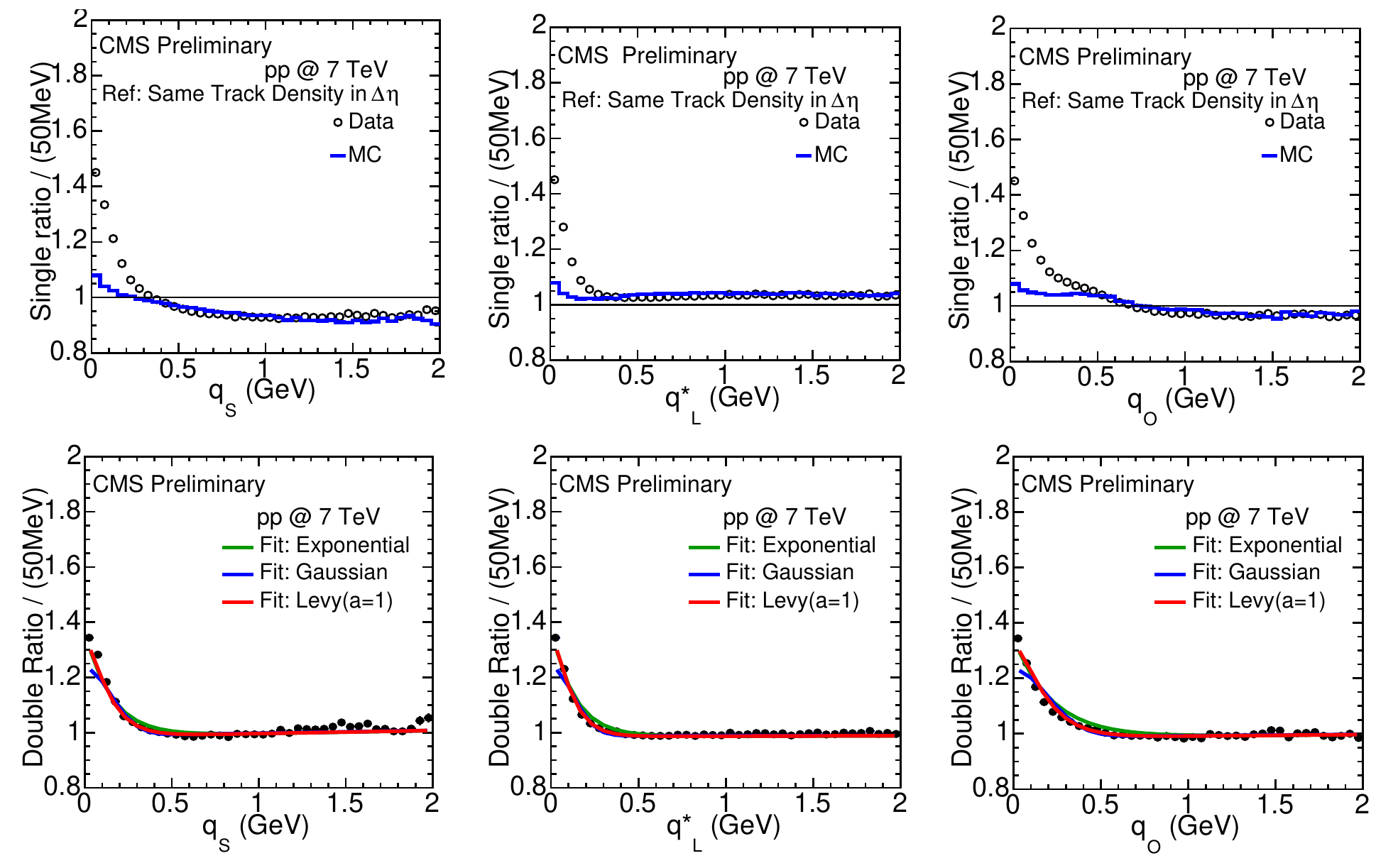}
    \caption{ The top panel shows the 1-D projections of the 3-D single ratios of data and Monte Carlo, in terms of the relative momentum components ($q_S$, $q^*_L$, $q_O$) analyzed in the LCMS, for pp collisions at 7 TeV; the bottom panel shows the corresponding double ratios. The Gaussian, the exponential and the L\'evy (with $a=1$) fit functions are shown superimposed to the data points. }
    \label{fig:3dqsqlqo_1d_projections_lcms-7tev}
  \end{center}
\end{figure}

The 3-D correlation function in terms of the variables  $(q_S, q^*_L, q_O)$ can be visualised through 2-D projections in terms of the  combinations  $(q_S, q^*_L)$, $(q^*_L, q_O)$, and $(q_O,q_S)$, 
with the complementary components 
within $|q_O|<0.05$ GeV, $|q_S|<0.05$ GeV, $|q^*_L|<0.05$ GeV, respectively, corresponding to the width of the first bins. 
The 1-D projections in the LCMS of the  
single and double ratios are shown in Fig.~\ref{fig:3dqsqlqo_1d_projections_lcms-7tev} for pp collisions at 7 TeV. The points represent  the data and the curves 
the exponential, {\sl stretched exponential} (L\'evy-type with $a=1$) and Gaussian fit functions. The fits are performed to the 3-D double ratios and then projected 
in the directions of $q_S, q^*_L, q_O$, similarly to the projections of the data points.

The values of the lengths of homogeneity in the Bertsch-Pratt parameterization for pp collisions at 7 TeV, obtained with the {\sl stretched} exponential fit in the LCMS, integrating over all $N_{ch}$ and $k_T$ ranges, are summarized in Table~\ref{tab:radii-3d-7tev},  together with the corresponding intercept fit parameter. 
Comparing the lengths of homogeneity in the 3-D case in the LCMS, from Table~\ref{tab:radii-3d-7tev}  
it is found that 
$R^*_L  \approx 1.5~ R_O $ fm and $R^*_L \approx 1.2~ R_S $ at 7 TeV. Therefore, the source seems to be more elongated along the longitudinal direction in the LCMS also in the 3-D case, with the relation among the lengths of homogeneity such as $R^*_L > R_S > R_O$.

\begin{table}[hbt]
\begin{center}
\caption{ 3-D fit parameters for pp collisions at $\sqrt{s}=7$ TeV in the LCMS }
\begin{tabular}{l|c}
\hline
\multicolumn{1} {c|} {$\sqrt{s}$ }   	& \multicolumn{1} {c} {7 TeV} \\  \hline
\multicolumn{1} {c|} {$\lambda$}       &   $ 0.568 \pm 0.002 $  (stat.) $ \pm~0.065 $ (syst.).    \\  \hline
\multicolumn{1} {c|} {$R_O$ (fm)}      &  $ 1.370 \pm 0.004 $ (stat.)  $ \pm~0.275$ (syst.)    \\ \hline
\multicolumn{1} {c|} {$R_S$ (fm)}      &   $ 1.784 \pm 0.004 $ (stat.)  $ \pm~ 0.275$ (syst.)   \\ \hline
\multicolumn{1} {c|} {$R^*_L$ (fm)}   &   $ 2.105 \pm 0.005 $ (stat.)  $ \pm~0.275$ (syst.)   \\ \hline
\end{tabular}
 \label{tab:radii-3d-7tev}
\end{center}
\end{table}

\begin{figure}[h]
  \begin{center}
    \includegraphics[width=1\textwidth]{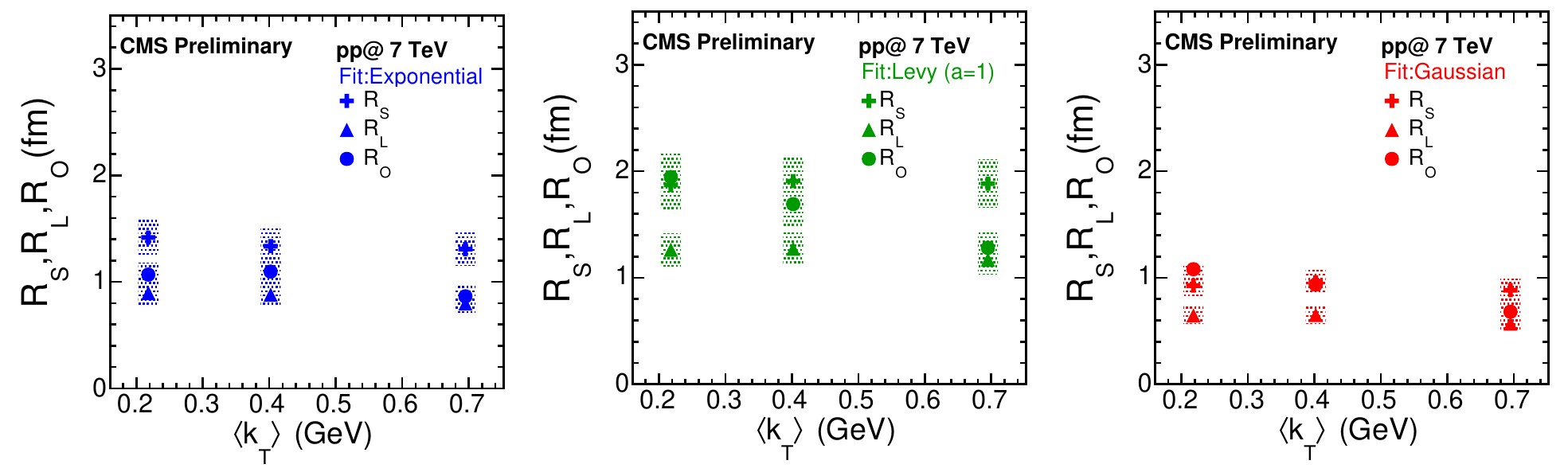}
    \includegraphics[width=1\textwidth]{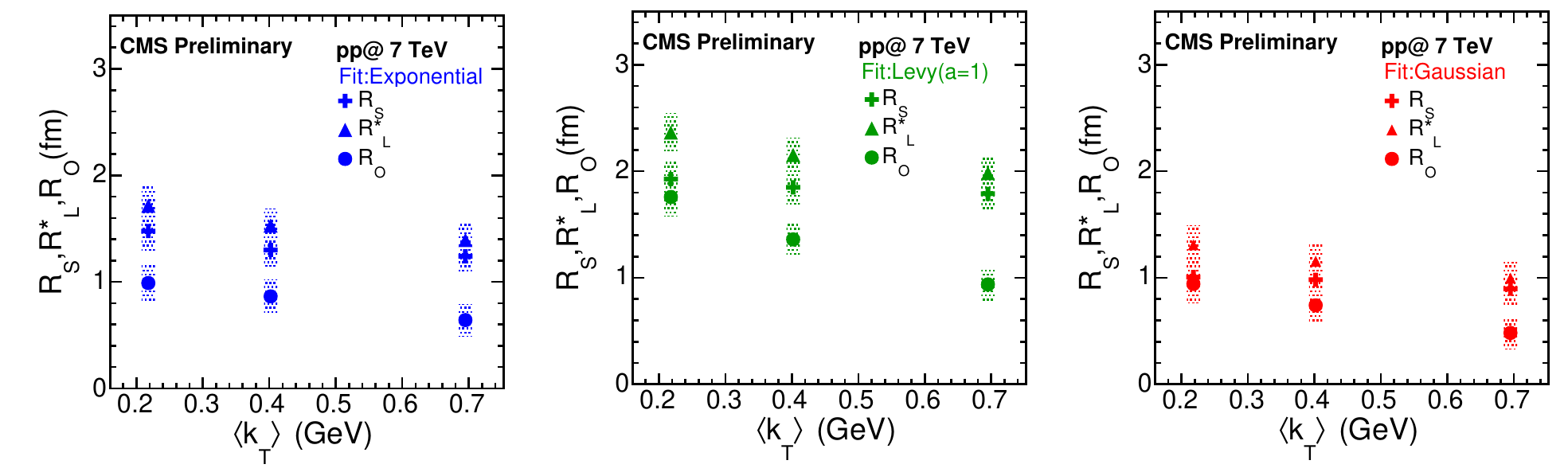}
    \caption{ The radius parameters in the CM frame (top) and in the LCMS (bottom), for pp collisions at 7 TeV, obtained from fits 
    to the double ratios with three different functions, are shown versus $\langle k_T \rangle$, integrated in $N_{ch}$. 
    The statistical uncertainties are indicated by error bars (in some cases, smaller than the marker's size), 
    whereas the systematic ones are indicated by shaded boxes. }
    \label{fig:fits_exp-gauss-levy-qsqlqo-kt_cm-lcms-7tev}
    \smallskip
    \includegraphics[width=1\textwidth]{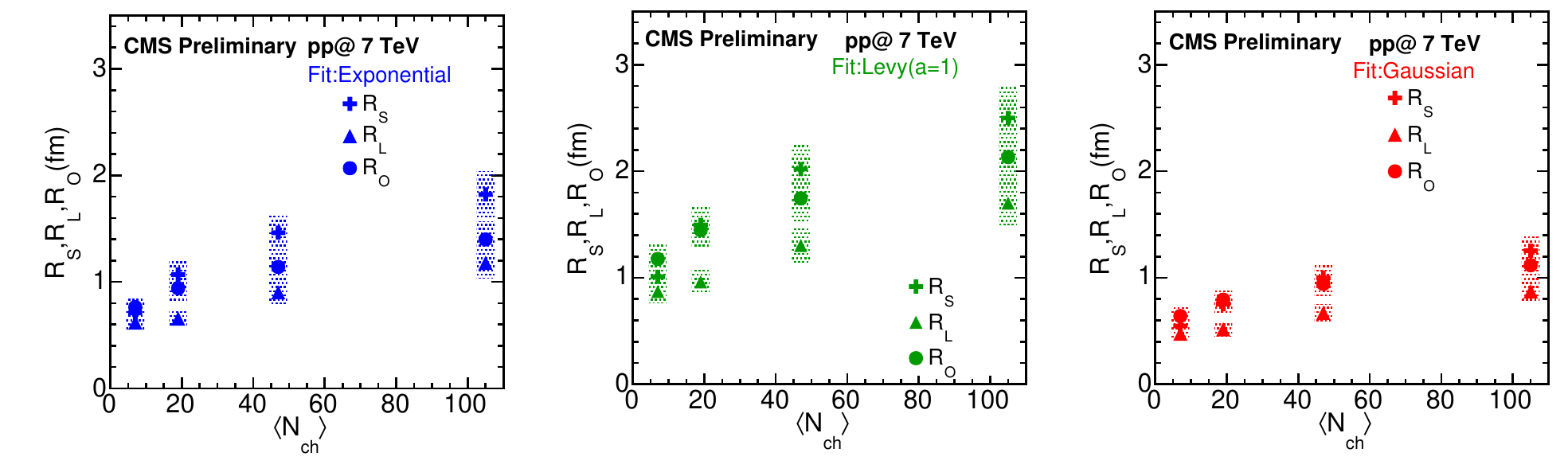}
    \includegraphics[width=1\textwidth]{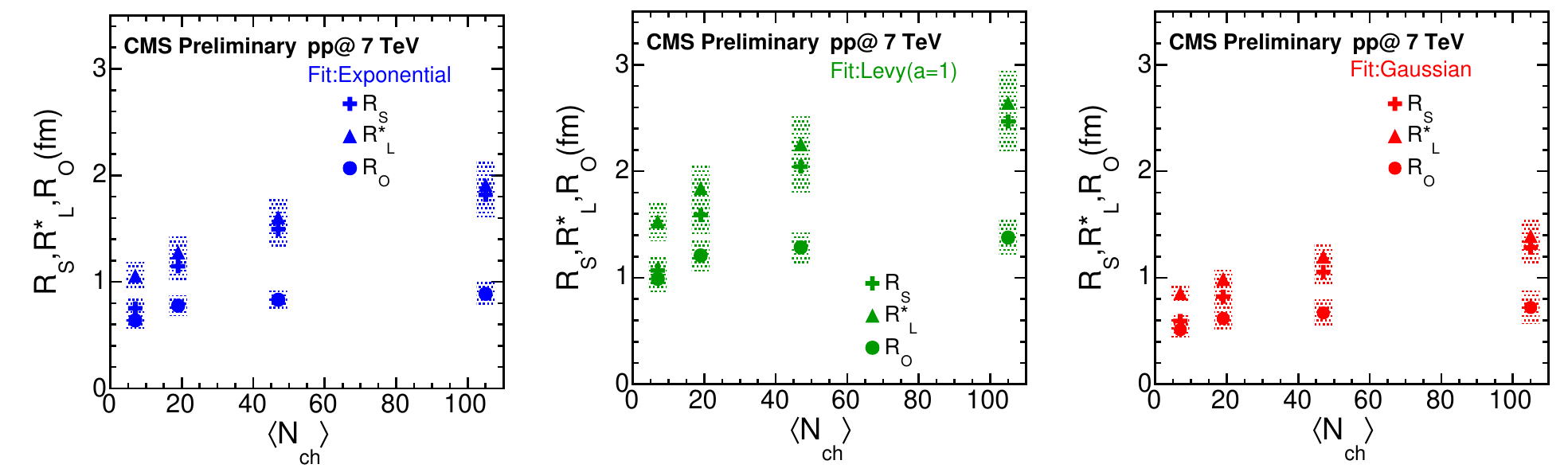}
    \caption{ Results of fits to the double ratios with three different fit functions in the CM frame (top) and in the LCMS (bottom)  for pp collisions at 7 TeV are shown as a function of $\langle N_{ch} \rangle$ (efficiency and acceptance corrected), integrated in $k_T$. 
    The statistical uncertainties are indicated by error bars (in some cases smaller than the marker's size), whereas the systematic 
    ones are indicated by the shaded boxes. }
    \label{fig:fits_exp-gauss-levy-qsqlqo-nch_cm-lcms-7tev}
  \end{center}
\end{figure}

The fits to the 3-D double ratios were also investigated in three $k_T$ bins 
 (integrating over all $N_{ch}$). The fit parameters were obtained with Gaussian, exponential and L\'evy (with $a=1$) fit functions  (when treated as a fit parameter $a$ returned values close to unity also in the 3-D case, both in the CM and in the LCMS). The results are compiled in Fig.~\ref{fig:fits_exp-gauss-levy-qsqlqo-kt_cm-lcms-7tev} for the data from pp collisions at 7 TeV, showing to depend noticeably on the type of fit used, the radius parameters being considerably larger in the case of the L\'evy-type function. The dependence on $k_T$, however, seems to be similar for the three fit functions. The $R_S$ fit values are among the largest (except for the Gaussian fit) and seem insensitive to $k_T$, in both 
frames. Also $R_L$ seems to be insensitive to $k_T$, and is the smallest radius parameter in the CM frame. However, it  shows opposite behavior in the LCMS, where its decrease with increasing $k_T$ is more pronounced, also attaining the largest values of the three radius parameters, suggesting an effect related to the Lorentz boost in the longitudinal direction, as in the 2-D case. The $R_O$ fit values are slightly smaller in the LCMS as compared to the CM frame, and decrease moderately with increasing $k_T$. Its dependence on $k_T$ is similar in both frames, although it has a slightly steeper decrease with increasing $k_T $ in the LCMS. 

The fits to the double ratios were also studied in four $N_{ch}$ bins (integrating over all $k_T$). 
The corresponding results are shown as a function of $\langle N_{ch} \rangle$ (efficiency and acceptance  corrected) in Fig.~\ref{fig:fits_exp-gauss-levy-qsqlqo-nch_cm-lcms-7tev}, obtained both in the CM frame (top) and in  the LCMS (bottom). 
A clear behavior can be seen, common to all fit functions and in all directions of the relative momentum components: the fit radius parameters $R_S$, $R_L$ and $R_O$ increase with increasing average multiplicity, indicating an increase in the lengths of homogeneity with $N_{ch}$, similar to what was seen in the 1-D and 2-D cases. 

\begin{figure}[hbt]
  \begin{center}
    \includegraphics[width=0.35\textwidth]{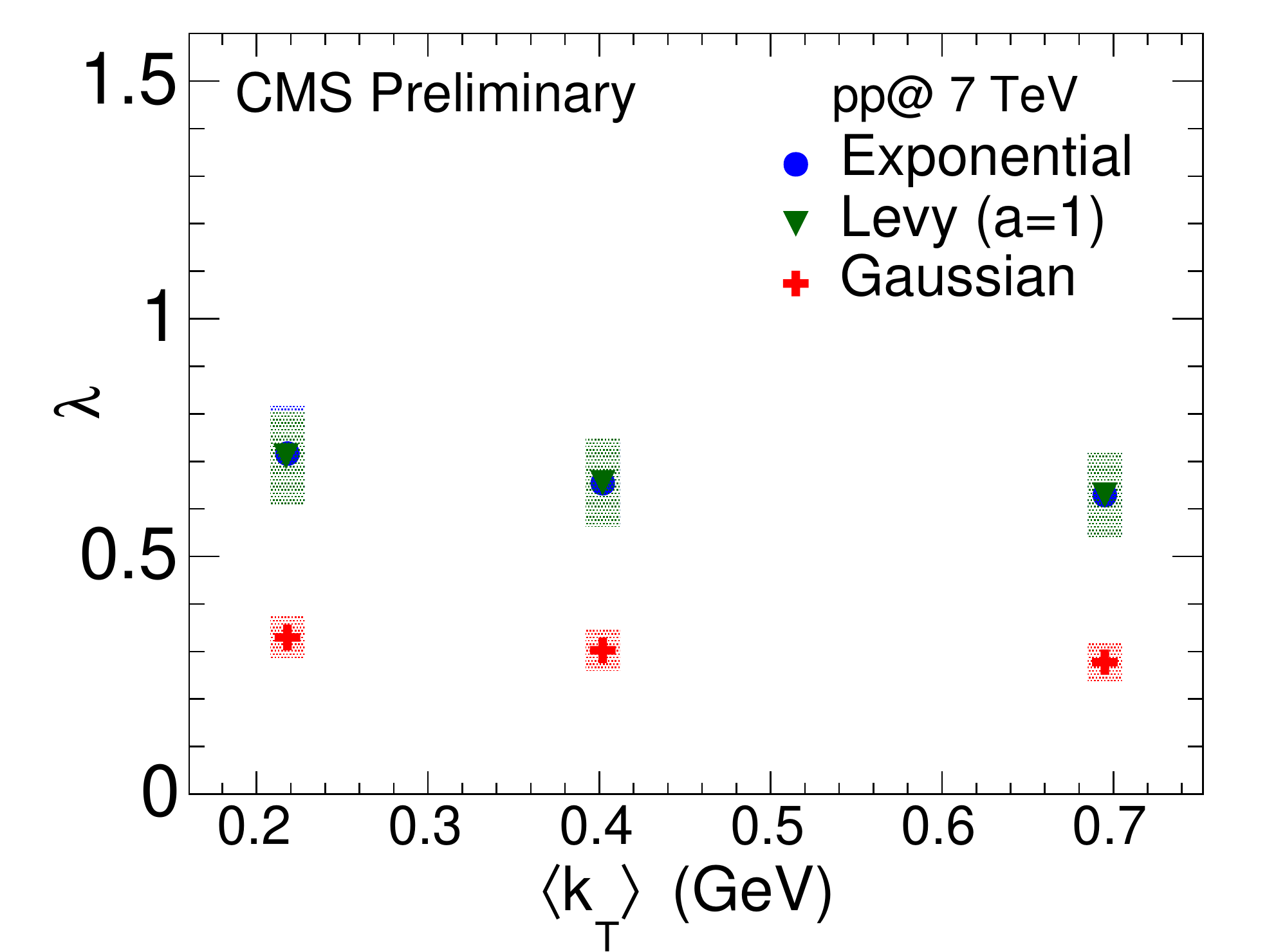}
    \includegraphics[width=0.35\textwidth]{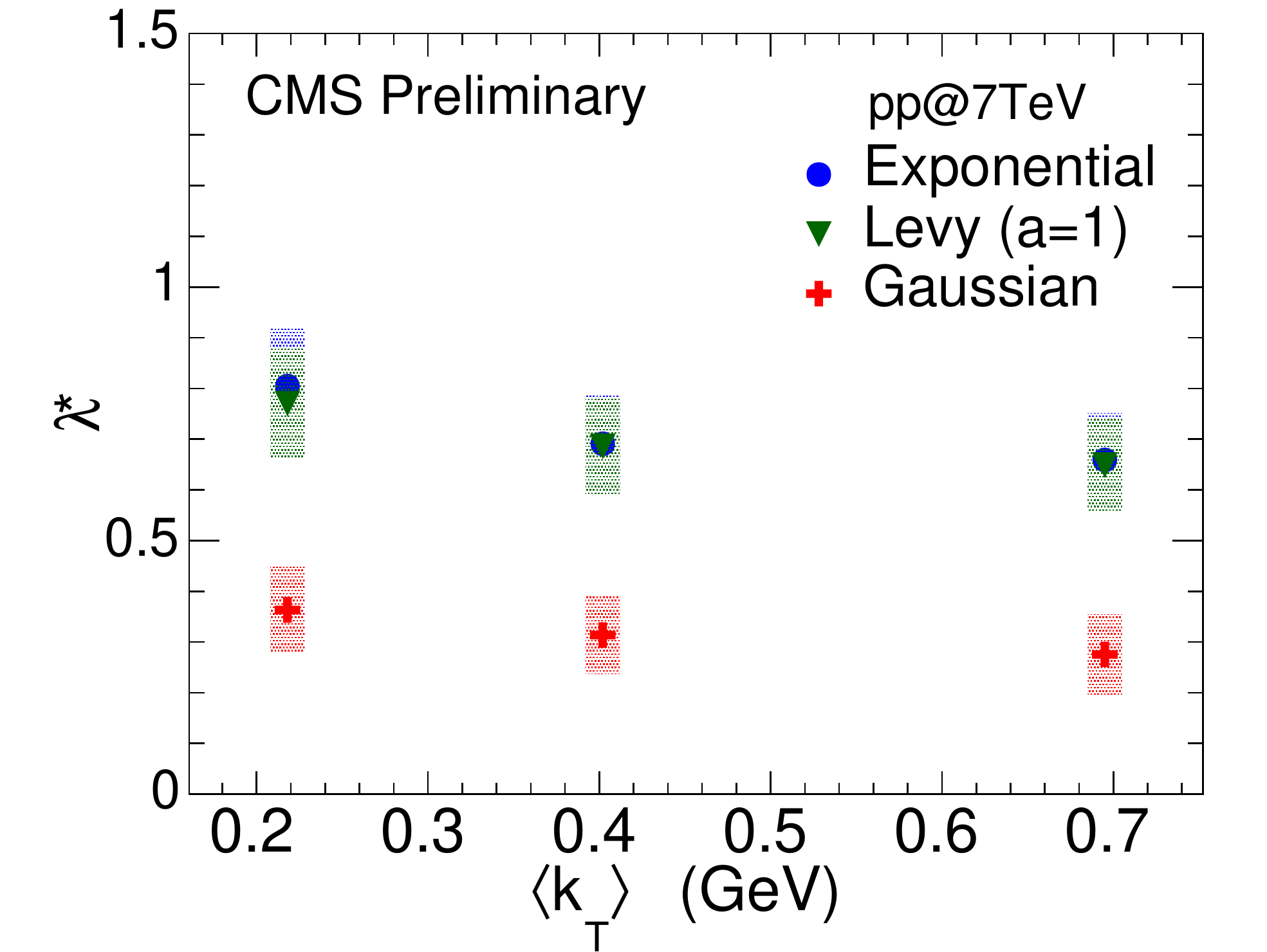}
    \includegraphics[width=0.35\textwidth]{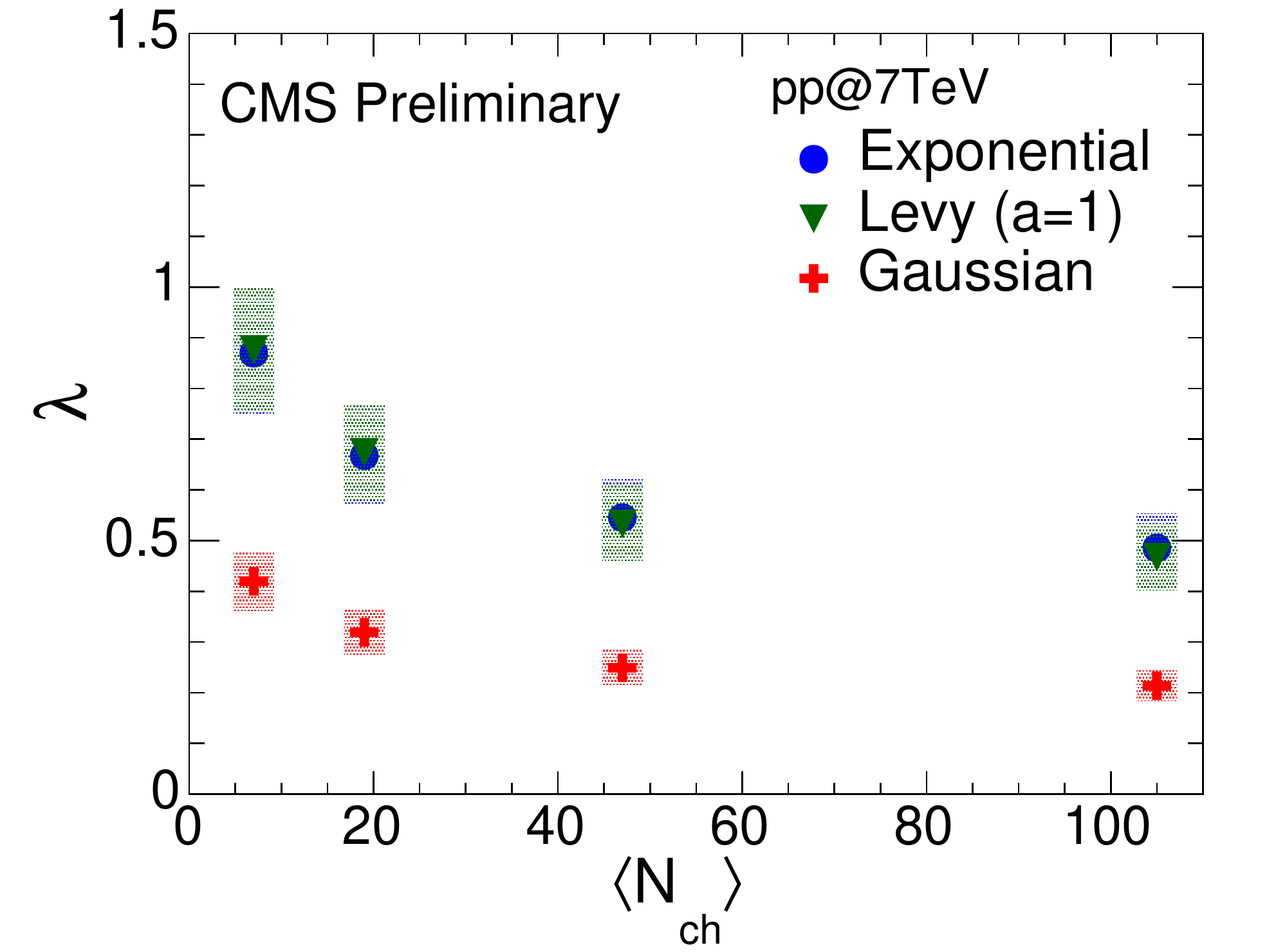}
    \includegraphics[width=0.35\textwidth]{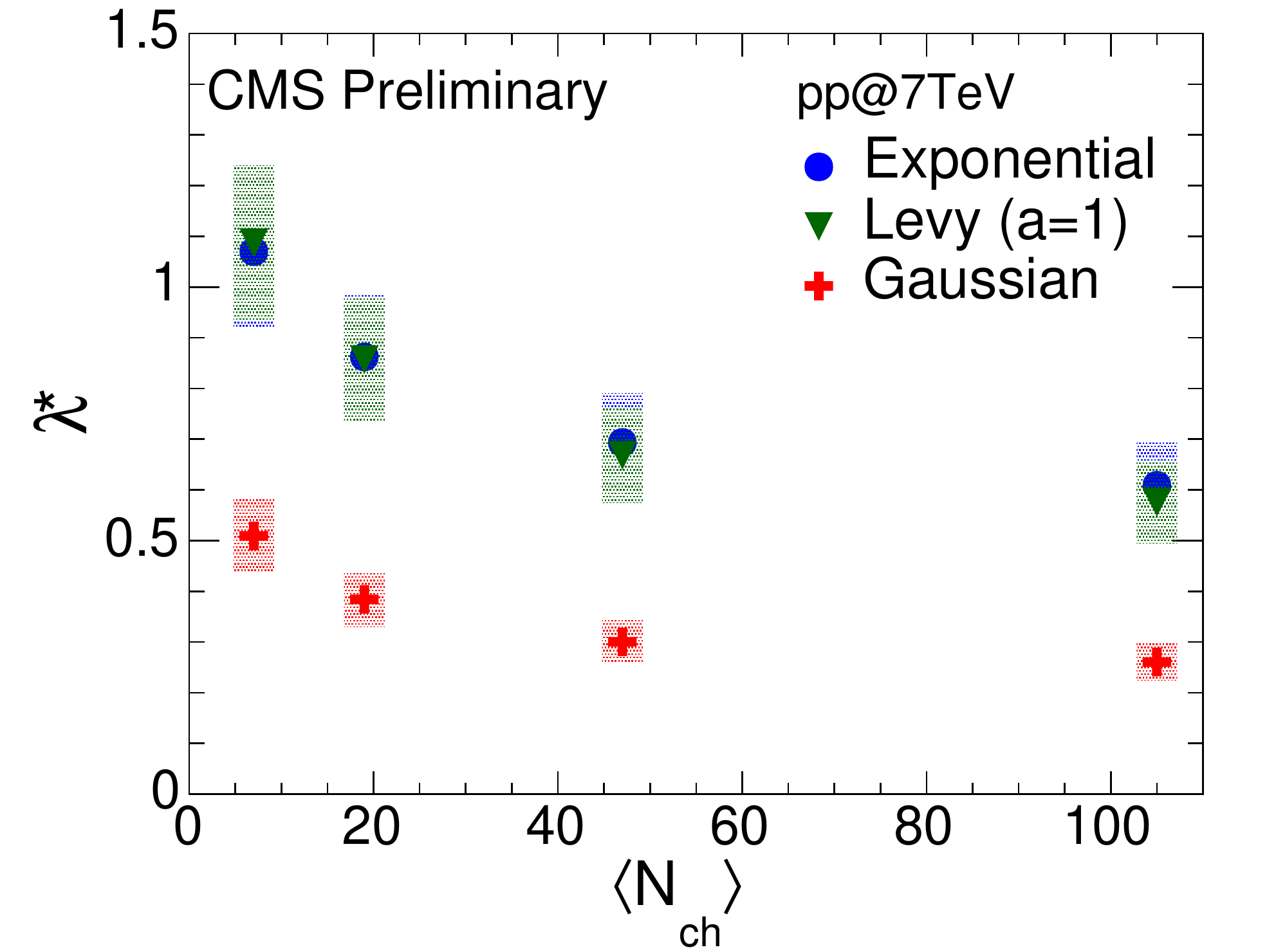}
    \caption{ Results corresponding to the intercept parameter $\lambda$ in the CM frame (left panel) and 
    in the LCMS (right panel) are shown on top as a function of $\langle k_T \rangle$ (integrated in $N_{ch}$), and on the bottom as a function of 
    $\langle <N_{ch}> \rangle$  (integrated in $k_T$). The values of $N_{ch}$ were corrected for efficiency and acceptance. 
    Statistical uncertainties are indicated by error bars (in some 
    cases smaller than the marker's size), whereas systematic ones are indicated by shaded boxes. }
    \label{fig:lambda-3d-kt-nch-cm-lcms}
   \includegraphics[width=0.32\textwidth]{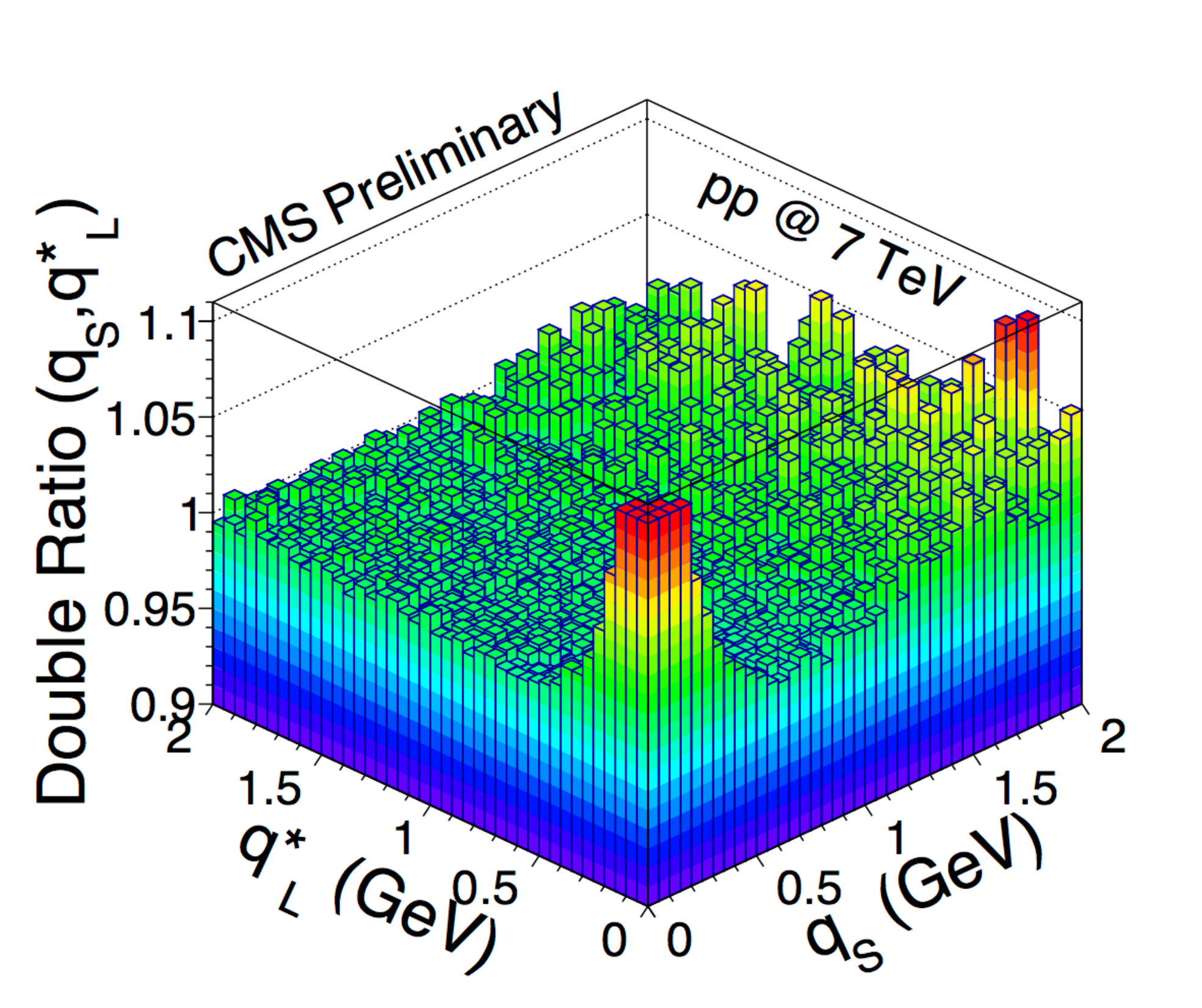}
   \includegraphics[width=0.32\textwidth]{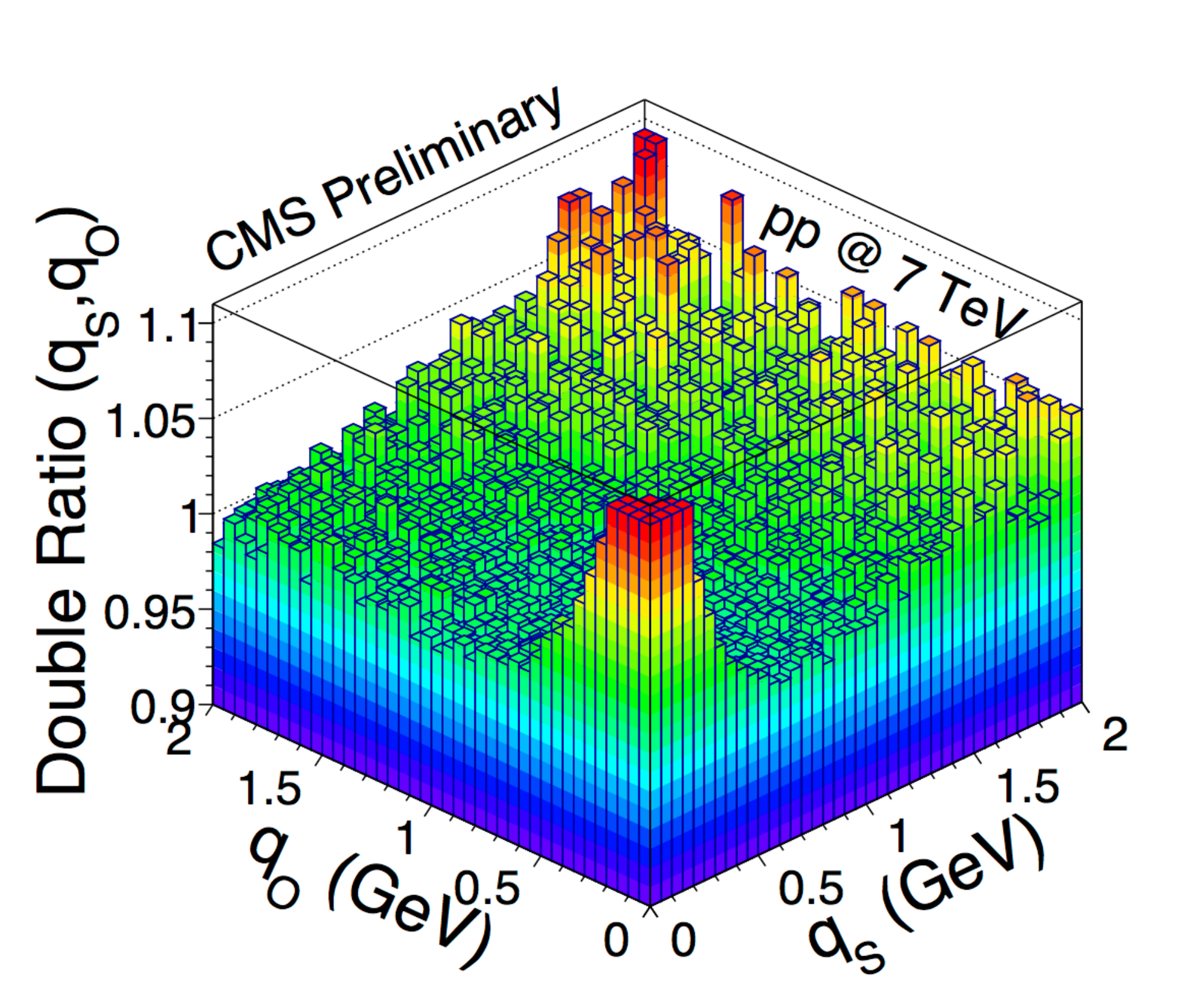}
   \includegraphics[width=0.32\textwidth]{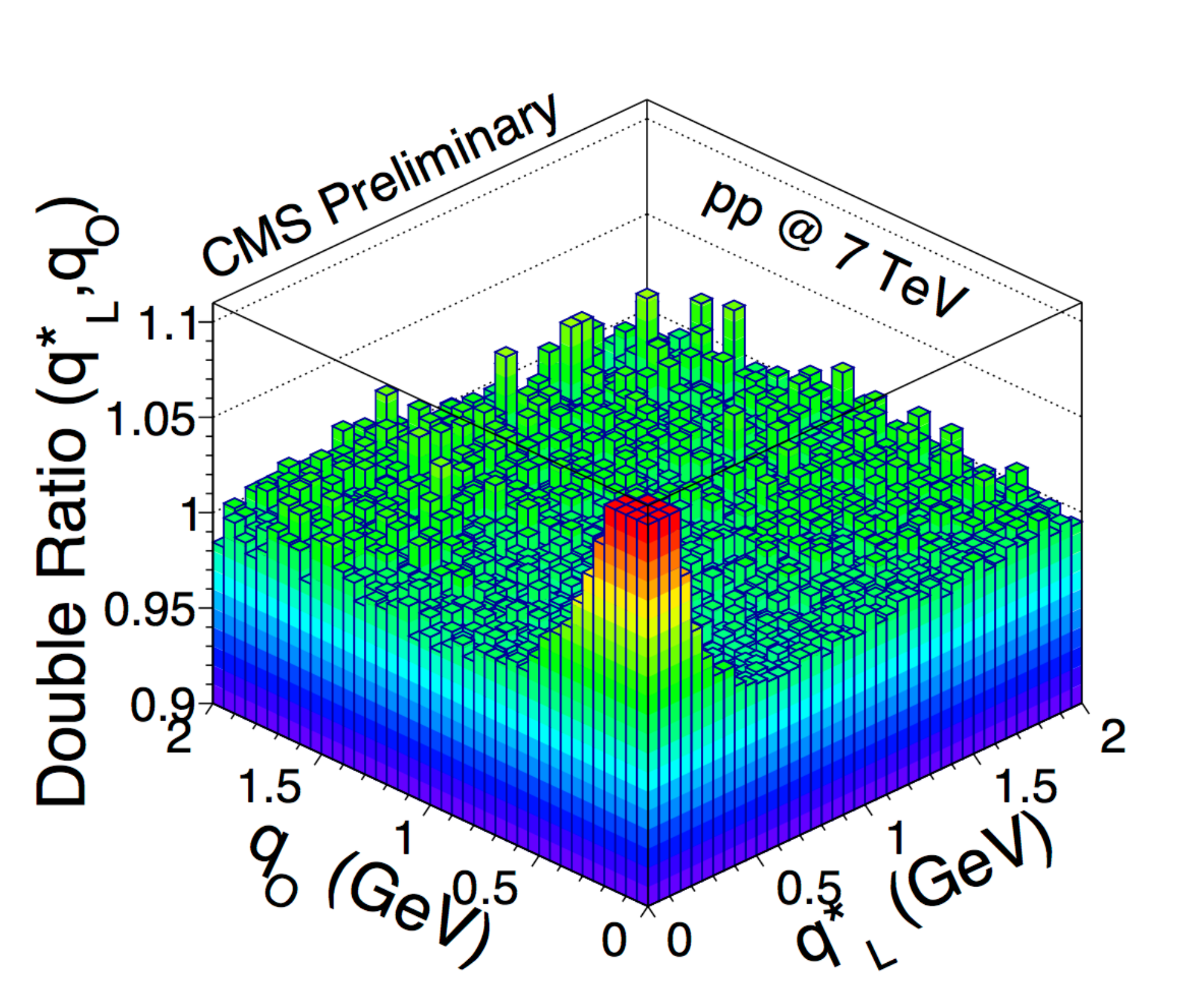}
   \includegraphics[width=1\textwidth]{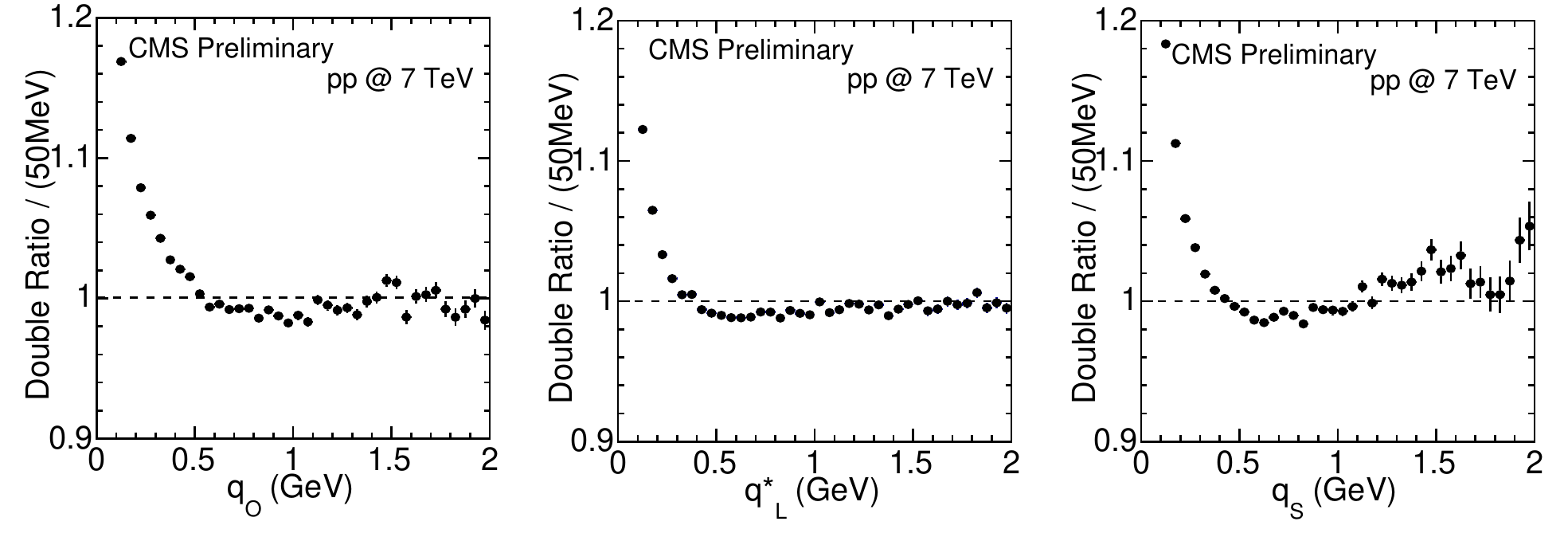}
  \caption{ The results for the 3-D double ratios obtained in the LCMS are shown, the upper panel corresponding 
  to the 2-D projections in ($q^*_L,q_S$), ($q_O,q_S$), and ($q_O,q^*_L$), integrated in $N_{ch}$ and $k_T$, with 
  $|q_O| < 0.05$ GeV, $|q^*_L| < 0.05$ GeV and $|q_S| < 0.05$ GeV, respectively. The bottom panel shows 1-D 
  projections of the same data, with the complementary two variables constrained to be within the first bin 
  (i.e., $|q_{i,j}| < 0.05$ GeV).
}
   \label{fig:3d-dip-structure-lcms} 
   \end{center}
\end{figure}

The intercept parameter $\lambda$ was also studied as a function of $k_T$ and $N_{ch}$, both in the CM frame and in the LCMS. The corresponding results are shown in Fig.~\ref{fig:lambda-3d-kt-nch-cm-lcms}. The values of $\langle N_{ch} \rangle$ shown in the plots were corrected for efficiency and acceptance. A moderate decrease with increasing $k_T$ is observed. As a function of increasing $N_{ch}$, $\lambda$ first decreases and then seems to saturate.

In the 2-D and 3-D cases, the values of the longitudinal radius fit parameters 
coincide within the experimental uncertainties  \cite{fsq-13-002-pas}, as expected, since both correspond to the length of homogeneity in the beam direction. From Figures \ref{fig:fits_exp-gauss-levy-qsqlqo-kt_cm-lcms-7tev} and \ref{fig:fits_exp-gauss-levy-qsqlqo-nch_cm-lcms-7tev} it can seen that  there is an approximate scaling of $R_L$ ($R^{*}_L$) and $R_T$ with $N_{ch}$, when comparing the results at 2.76 TeV and at 7 TeV.  

In Fig.~\ref{fig:3d-dip-structure-lcms} the 3-D results for the double ratios in the LCMS (integrated in $N_{ch}$ and $k_T$) are shown
as 2-D projections in terms of pairs of $q_S, q_L, q_O$ (the complementary one within $| q_i |< 0.05$ GeV).  The plots are zoomed along the correlation function axis, cutting values above 1.2. The corresponding 1-D projections along  variable $q_S, q_L,$ and $q_O$ (other two variables within $| q_{i,j} |< 0.05$ GeV) are also shown.

\clearpage

\section{Summary and conclusions} 
\label{sec:summ_conclusions}

The analysis discussed here extends the 1-D results reported in the two BEC publications in \cite{cms-hbt-1st,cms-hbt-2nd} by measuring the correlation functions in terms of different components of the pair relative momentum, as is usually studied by other experiments. This allows to investigate the extension of the source accessible to the correlation technique in different directions. Two main projections are considered: in two-dimensions (2-D), the femtoscopic correlation is investigated as a function of the variables $q_L$ and $q_T$, and  
in three-dimensions (3-D), as function of $q_L, q_S, q_O$, being $q_L$ the same as in the 2-D case. 
For achieving this purpose, minimum bias events produced in proton-proton collisions at 2.76 and 7 TeV (full data sample) are scrutinised in detail, as if by means of a magnifying lens. At 7 TeV, the full data sample is used and the corresponding results are compared with the ones at the same energy recorded during the commissioning run at the LHC in 2010. This analysis 
also extends the measurements of the BEC correlations to the full minimum bias sample from pp collisions at 2.76 TeV collected in 2013, which is a very important baseline for the measurements of this second order interferometry in PbPb collisions at the same energy per nucleon. In 1-D are the results from both energies are  compared with the ones in Ref. \cite{cms-hbt-1st,cms-hbt-2nd} at lower energies, as well as at $\sqrt{s}=7$~TeV, recorded during the commissioning run at the LHC in 2010. In particular, comparisons showed that $R_\textrm{inv}$ steadly increases with the charged multiplicity proportionally to $N_{ch}^{1/3}$.  

The measurements were performed both in the collision center-of-mass (CM) frame and in the Local Co-Moving System (LCMS), where the average longitudinal momentum of the pair is zero. In the 2-D case, for integrated values of $N_{ch}$ and $k_T$, the lengths of homogeneity in the LCMS  suggest that the source is elongated along the beam direction, i.e., $R^*_L > R_T$. In the 3-D case, it was found that $R^*_L > R_S > R_O$. 
 In addition, it can be observed that the fit values for the longitudinal radius parameter, $R_L$ are consistent in 2-D and in 3-D cases, as should be expected, since they correspond to the length of homogeneity in the longitudinal direction in both cases. This conclusion is attained with respect to the $R_L$ fit parameter in the CM frame, as well as $R^*_L$ in the LCMS \cite{fsq-13-002-pas}. 

The anticorrelation observed in 1-D and reported in Ref.~\cite{cms-hbt-2nd}, was also observed in minimum bias pp collisions at 2.76 TeV and further investigated here with the full statistics at 7 TeV. 
These new 1-D results are also compared with those in Ref.~\cite{cms-hbt-2nd}, verifying a consistent behavior  both for the invariant radius parameter and for the dip's depth measurements at 2.76 and 7 TeV. Such comparisons showed that the dip's depth decreases with increasing $N_{ch}$.

\end{document}